\pdfoutput=1
\documentclass[11pt,letterpaper]{article}
\PassOptionsToPackage{dvipsnames}{xcolor}

\usepackage[T1]{fontenc}
\usepackage{lmodern}
\usepackage{amsmath,amssymb,amsthm,mathtools}
\usepackage[letterpaper,margin=1in]{geometry}
\usepackage[ruled,vlined]{algorithm2e}
\usepackage{microtype}
\usepackage{xcolor}
\usepackage{tikz}
\usepackage[font=small,labelfont=bf]{caption}
\usetikzlibrary{calc,arrows.meta}
\usepackage[colorlinks=true,citecolor=blue,linkcolor=blue,urlcolor=blue]{hyperref}
\usepackage[nameinlink,noabbrev]{cleveref}
\usepackage{aliascnt}
\usepackage{enumitem}
\usepackage{booktabs,multirow}

\newcommand{\OPT}{\operatorname{OPT}}
\newcommand{\OPTLP}{\OPT_{\mathrm{LP}}}

\DeclareRobustCommand{\md}{\mathsf{md}}

\theoremstyle{plain}
\newtheorem{theorem}{Theorem}[section]
\newaliascnt{lemma}{theorem}
\newtheorem{lemma}[lemma]{Lemma}
\aliascntresetthe{lemma}
\newaliascnt{corollary}{theorem}
\newtheorem{corollary}[corollary]{Corollary}
\aliascntresetthe{corollary}
\theoremstyle{definition}
\newaliascnt{definition}{theorem}
\newtheorem{definition}[definition]{Definition}
\aliascntresetthe{definition}
\theoremstyle{remark}
\newaliascnt{remark}{theorem}
\newtheorem{remark}[remark]{Remark}
\aliascntresetthe{remark}
\crefname{lemma}{Lemma}{Lemmas}
\crefname{corollary}{Corollary}{Corollaries}
\crefname{definition}{Definition}{Definitions}
\crefname{remark}{Remark}{Remarks}

\title{Minimum Sum-of-Radii Clustering with Strong and Weak Oracles}
\author{
 Zachary Friggstad\thanks{University of Alberta. Email: \texttt{zacharyf@ualberta.ca}. Research supported by an NSERC Discovery Grant.}\\
 \and
 Mahya Jamshidian\thanks{University of Alberta. Email: \texttt{mjamshidian@ualberta.ca}.}\\
 \and
 Ali Vakilian\thanks{Rutgers University. Email: \texttt{a.vakilian@rutgers.edu}. Supported in part by NSF CAREER award CAIF-2638918.}\\
}
\date{}

\makeatletter
\def\input@path{{arxiv/}}
\makeatother

\begin{document}
\maketitle
\thispagestyle{empty}

\begin{abstract}
In the minimum sum of radii (MSR) problem, the input is a set of $n$ points in a metric space and a parameter $k$, and the goal is to cover the points by at most $k$ balls, each centered at an input point, while minimizing the sum of their radii. We study MSR in the weak–strong oracle model of Bateni, Dharangutte, Jayaram, and Wang (COLT’24), where the metric is unknown and is accessed through two oracles. The answers of the weak oracle are given to us for free: for every pair of points, we receive a distance value that, with probability $\delta$, is arbitrarily corrupted. Exact distances are available only through costly strong queries, where a strong query on a point reveals its exact distances to all previously strongly queried points, and the goal is to minimize the number of strong queries.

For every constant $\delta < 1/2$, we give the first polynomial-time
algorithm that makes $O(k \log n \log\log n)$ strong point queries and,
with high probability, returns an $O(1)$-approximation for MSR; the
previously best algorithm required running time and a number of strong
queries exponential in $k$. This is within an $O(\log\log n)$ factor of
a lower bound that we prove: any constant-factor approximation that
succeeds with high probability on every instance must make
$\Omega(k \log n)$ strong queries in expectation on some instance. The same $\Omega(k \log n)$ lower bound holds for $k$-center, where it matches the known upper bound up to an $O(\log\log n)$ factor.

An optimal MSR cluster can be small both in size and in radius, so a uniform sample of strongly queried points may miss it entirely. Instead, our algorithm uses a packing LP over the balls certified so far to decide which points to query next. The analysis holds against an adversary who knows which pairs are corrupted.
\end{abstract}

\setcounter{tocdepth}{2}
\newpage
\setcounter{page}{1}
\section{Introduction}\label{sec:intro}

A clustering algorithm takes a set of points with a notion of distance. It groups the points so that each group holds points that are close to each other.
Clustering is a fundamental problem in data analysis and machine learning.
It appears wherever data has to be summarized, compressed or searched, from biology to computer vision.
The usual way to make the task precise is to choose $k$ centres and to assign every point to one of them.
The cost of the solution then depends on the distances from the points to their centres.
Under $k$-center the cost is the largest such distance.
This problem has a $2$-approximation, and no better factor is possible~\cite{gonzalez1985clustering,hochbaum1985best}.

However, $k$-center minimizes only the largest cluster radius.
Once this radius is fixed, reducing any of the smaller radii does not improve the objective.
If one cluster needs a large radius, we would still like the other radii to be as small as possible~\cite{buchem2024msr}.
This motivates the \emph{minimum sum of radii} problem (MSR): choose at most $k$ balls covering all points and minimize the sum of their radii~\cite{doddi2000approximation,charikar2004clustering}.

For MSR the approximation factor in polynomial time went from $3.504$~\cite{charikar2004clustering} to $3.389$~\cite{friggstad2022msr} and then to $3+\epsilon$~\cite{buchem2024msr}.
All of these algorithms assume that every pairwise distance is available exactly.

Often in practice, finding or calculating exact distances between objects are costly when computed from learned representations.
Obtaining an accurate embedding for every point can be expensive too; but once two embeddings are available, their distance can be computed without another model call. 
Then, one can make use of a cheaper model to provide distances for all pairs, while the expensive model is invoked only on selected points when necessary. 
The cheap model may come from a smaller network, a model trained on noisy data, or a previously stored similarity matrix. 
Related trade-offs between cheap and reliable information appear in active learning with two labelers~\cite{zhang2015weakstrong} and correlation clustering with two similarity oracles~\cite{silwal2023kwikbucks}.

Computation with an unreliable oracle is a classical topic.
How many noisy answers are needed to sort, to search or to cluster has been studied for decades, also for persistent noise, where a repeated query returns the same wrong answer~\cite{braverman2008noisysorting,mazumdar2017clustering,peng2021faulty}.
Almost all of this work has a single noisy oracle.
Much less is known when there are two oracles, a noisy cheap one and an exact expensive one, and the question is how many exact answers are enough.
Bateni, Dharangutte, Jayaram and Wang~\cite{bateni2024metric} started the formal study of this setting for metric optimization, and they called it the \emph{weak-strong oracle model}.

\begin{definition}\label{def:model}
An instance of the weak-strong oracle model is a metric $(X,d)$ on $n$ points and a corruption probability $\delta\in(0,\frac12)$.
The algorithm does not know $d$, and it accesses $d$ only through two oracles.
\begin{itemize}[leftmargin=*]
    \item A \emph{strong query} on a point $x$ returns a value $\mathrm{SO}(x)$. The value $\mathrm{SO}(x)$ gives no information on its own. Given $\mathrm{SO}(x)$ and $\mathrm{SO}(y)$, the algorithm can compute $d(x,y)$.
    \item A \emph{weak query} on a pair $\{x,y\}$ returns a value $\Tilde d(x,y)$. With probability $1-\delta$ we have $\Tilde d(x,y)=d(x,y)$, and otherwise $\Tilde d(x,y)$ is set arbitrarily, in which case we call the pair \emph{corrupted}.
\end{itemize}
The randomness is independent across pairs and is drawn exactly once, so repeating a weak query gives the same value.

The cost of an algorithm is the number of strong queries.
\end{definition}

We study the \emph{minimum sum of radii} problem (MSR) in this model.
Let $\OPT$ denote the minimum sum of radii.
MSR is NP-hard even in the shortest-path metrics of weighted planar graphs and in metrics of constant doubling dimension~\cite{gibson2010metric}.
It has an exact quasi-polynomial-time algorithm when the aspect ratio is polynomially bounded~\cite{gibson2010metric}, and exact polynomial-time algorithms for points in the plane~\cite{gibson2012planar} and for the shortest-path metrics of unweighted graphs when singleton clusters are not allowed~\cite{behsaz2015msr}.
Unlike $k$-center, MSR has no single distinguished radius.
An optimal solution may contain clusters at very different scales, and the cost is their sum.
So it is not clear how to find the useful clusters with a small number of strong queries.
Following~\cite{bateni2024metric}, a solution consists of at most $k$ balls together with an assignment of every point to a ball that contains it, so the algorithm returns the clustering and not only the centres.

\subsection{Our Results}

Our main result is a polynomial-time constant-factor approximation for MSR against the adversary of \Cref{sec:model}. As in~\cite{bateni2024metric}, we assume that a polynomial upper bound on the metric's aspect ratio is known.

\begin{theorem}\label{thm:intromain}
Let $\delta\in(0,\frac12)$ be a constant.
There is a polynomial-time algorithm that makes $O(k\log n\log\log n/(\frac12-\delta)^3)$ strong point queries, and with high probability returns at most $k$ balls and an assignment of every point to a returned ball that contains it, of total radius $O(\OPT)$.
\end{theorem}

The approximation factor does not depend on $\delta$.
In the main text we fix $\delta=\frac13$ and prove \Cref{thm:intromain} for this value ( $O(k\log n\log\log n)$ many strong queries).
\Cref{app:delta} proves it for every constant $\delta\in(0,\frac12)$ (\Cref{thm:generaldelta}).
\Cref{app:factor} computes the approximation factor, which is less than $270$ (\Cref{thm:factor}).

\Cref{thm:intromain} is the first polynomial-time constant-factor approximation for MSR in the model.
The upper bound is within a factor $O(\log\log n)$ of our $\Omega(k\log n)$ lower bound for algorithms that succeed with high probability.

Our lower bound applies to any algorithm that returns a constant-factor approximation for MSR, including algorithms that inspect all weak distances and output centres they have not strongly queried. 
\begin{theorem}\label{thm:introlb}
For every fixed $\beta>0$ and $3\le k\le n/\log^2 n$, an algorithm that returns a constant-factor approximation for MSR with probability at least $1-n^{-\beta}$ on every instance must make $\Omega(k\log n)$ strong queries in expectation on some instance.
The same bound holds for $k$-center even when only centres are required.
\end{theorem}

\Cref{thm:whp,cor:kcenter} prove this.
Even when the algorithm only needs to succeed with constant probability, meaning at least a fixed $\rho>0$ independent of $n$ on every instance, the same expected-query lower bound holds when $n=k^2$ (\Cref{thm:constant}).
\Cref{tab:compare} summarizes the bounds.
Our earlier MSR algorithm~\cite{altsubmission} has running time and strong queries exponential in $k$, and our lower bound there holds only for the {\em row-local} algorithms of $k$-center. 
An algorithm is row-local if each point’s cluster assignment depends only on its weak distance profile to the strongly queried points. All known algorithms for clustering problems~\cite{bateni2024metric,pradhan2026noisy,altsubmission}, including those in this paper, are row-local.

\begin{table*}[t]
\centering
\small

\begin{tabular*}{\textwidth}{@{\extracolsep{\fill}}lclcl@{}}
\toprule
\multicolumn{5}{@{}l}{\textbf{Upper bounds}} \\
\midrule
Objective & Approx.\ factor & Strong queries & Query type & Reference \\
\midrule
\multirow{3}{*}{$k$-center}
  & $14+\epsilon$
  & $O(k\log^2 n\log(\epsilon^{-1}\log n))$
  & Point
  & \cite{bateni2024metric} \\
  & $6+\epsilon$
  & $O(k^3\log^2 n\log(\epsilon^{-1}\log n))$
  & Edge
  & \cite{pradhan2026noisy} \\
  & $6+\epsilon$
  & $O(k\log n\log(\epsilon^{-1}\log n))$
  & Point
  & \cite{altsubmission} \\
\midrule
$k$-median, $k$-means
  & $O(1)$
  & $O(k\log^2 n)$
  & Point
  & \cite{bateni2024metric} \\
$k$-means
  & $(40+\epsilon)^\dagger$
  & $O(k^2\log^2 n/\epsilon^8)$
  & Edge
  & \cite{pradhan2026noisy} \\
\midrule
\multirow{2}{*}{MSR}
  & $40$
  & $2^{O(k\log k)}\cdot O(k\log n\log\log n)$
  & Point
  & \cite{altsubmission} \\
  & $O(1)$
  & $O(k\log n\log\log n)$
  & Point
  & {\bf \Cref{thm:intromain}} \\
\bottomrule
\end{tabular*}
\vspace{0.8em}

\begin{tabular*}{\textwidth}{@{\extracolsep{\fill}}llcl@{}}
\toprule
\multicolumn{4}{@{}l}{\textbf{Lower bounds \scriptsize{(for $O(1)$-approximation algorithms)}}} \\
\midrule
Objective & Setting & Strong (point) queries & Reference \\
\midrule
$k$-center, $k$-median, $k$-means
  & General 
  & $\Omega(k)$
  & \cite{bateni2024metric} \\
$k$-center
  & Row-local algorithms
  & $\Omega(k\log(n/k))$
  & \cite{altsubmission} \\
MSR, $k$-center
  & General
  & $\Omega(k\log n)$
  & {\bf \Cref{thm:introlb}} \\
\bottomrule
\end{tabular*}

\caption{Comparison of approximation guarantees and strong-query
complexities in the weak--strong oracle model.
Point and edge queries follow the conventions of~\cite{bateni2024metric},
and each upper bound uses the convention of its corresponding work.
The $k$-means result marked $\dagger$ uses
$O(k\log n/\epsilon^4)$ centers.}
\label{tab:compare}
\end{table*}

\subsection{Technical overview}

For each positive guess $\lambda$ of $\OPT/k$, the algorithm constructs candidates for potential balls (clusters) defined with a strong-queried centre and a radius, rounds them, repairs their assignments, and compresses them to at most $k$ balls (\Cref{sec:overview}). A binary search makes $O(\log\log n)$ such multi-step executions.
Throughout, $h=\Theta(\log n)$ is the number of witnesses used in a distance estimate.

\begin{itemize}[leftmargin=*]
\item \emph{Sampling guided by a packing LP:}
For $k$-center, one can guess the optimal radius $R$, remove the points within $O(R)$ of the samples and sample again. 
MSR has no single radius, so we do not know which points to remove, and uniform sampling may miss an optimum cluster that is small in size and in radius.
Trying each radius scale in turn does not fix this: with one cluster of radius $R$ and $k-1$ clusters of radius close to $0$, large scales can cost $\Theta(kR)$ while $\OPT$ is about $R$, and small scales do not remove the large cluster.
Instead, a packing LP, related to the dual of the Lagrangian relaxation of the standard MSR covering LP, assigns weight to points not yet accounted for by candidate balls (not close to any established centre, i.e. candidate ball). We sample in proportion to this weight.

\item \emph{Repair:}
Then the algorithm assigns the unqueried points to one of $O(k)$ balls using median weak distances to a witness set, a set of $h$ strongly queried points, as the core of some candidate ball.
An incorrect assignment can occur only for points with at least $h/2$ corrupted pairs to some witness set.

So we test every point again against new witnesses, drawn uniformly from candidate balls.
If the candidate ball has enough points that not too many of them can be corrupted, then new samples makes it trustworhty; else, we strong query all points inside smaller candidate balls.

\item \emph{Lower bound:}
We take $\Theta(k)$ small clusters of $\Theta(\log n)$ points each, and each may contain one outlier, a point far from every other point that the weak distances may not show.
Every approximate solution must choose every outlier as a centre to prevent an unbounded bump in the approximation factor, so the algorithm has to find these outliers with strong queries.
\end{itemize}

\subsection{Related work}

Bateni et al.~\cite{bateni2024metric} gave the first algorithms in the model, and they allow an adversary to choose the values of the corrupted pairs (\Cref{sec:model}).
For $k$-center they guess the optimal radius $R$ up to a factor $1+\epsilon$, draw uniform strong samples in $O(\log n)$ rounds, and keep every point whose median weak distance to the samples around a centre is $O(R)$, which gives a $(14+\epsilon)$-approximation with $O(k\log^2n\log(\epsilon^{-1}\log n))$ strong point queries.
For $k$-median and $k$-means they build a coreset by successive sampling with a proxy distance, which gives an $O(1)$-approximation with $O(k\log^2n)$ strong point queries.
They also consider strong \emph{edge} queries, which reveal one exact distance each. They prove that any algorithm achieving a fixed bounded approximation factor with probability at least $1/2$ for any of these three objectives needs $\Omega(k)$ strong point queries or $\Omega(k^2)$ strong edge queries.
Their construction can be adapted to establish an $\Omega(k)$ strong-point lower bound for MSR as well. For every fixed approximation factor, our lower bound shows that an MSR algorithm that succeeds with high probability requires $\Omega(k\log n)$ strong point queries in expectation, in the parameter range of \Cref{thm:whp}.

Pradhan, Bhattacharya and Jaiswal~\cite{pradhan2026noisy} count strong edge queries.
They run $k$-means++ on estimated distances, which gives a bicriteria $(40+\epsilon)$-approximation for $k$-means with $O(k\log n/\epsilon^4)$ centres and $O(k^2\log^2n/\epsilon^8)$ strong edge queries, and a ball-carving algorithm for $k$-center with factor $(6+\epsilon)$ and $O(k^3\log^2n\log(\epsilon^{-1}\log n))$ strong edge queries.
They also proved that $\Omega(nk/(1-2\delta)^2)$ weak edge queries are necessary for $k$-means.

Bateni et al.~\cite{bateni2024metric} gave a tight $O(\sqrt{\log n})$-approximation for the minimum spanning tree (MST) problem without strong queries, assuming the corrupted distances still form a metric. However, no approximation guarantee was known for the general setting, where corrupted distances may violate the triangle inequality. Recently, Anand et al.~\cite{anand2026mst} resolved this open question by obtaining a tight $O(\log n)$-approximation using only corrupted distances, for corruption probability $p<1/4$.
Braverman et al.~\cite{braverman2025fair} build coresets for fair clustering in the model.
No polynomial-time constant-factor approximation for the sum of radii was known in this model.
A parallel line of work replaces the weak distances by a noisy oracle that only compares two distances, together with a few exact distance queries~\cite{galhotra2024comparison,raychaudhury2025comparison}, or with no exact queries at all~\cite{raychaudhury2026comparison}.
The last of these lists the sum of radii as an open direction.
The idea of a cheap noisy signal next to an expensive exact one is older than these models.
Zhang and Chaudhuri~\cite{zhang2015weakstrong} study active learning with a weak and a strong labeler, and Silwal et al.~\cite{silwal2023kwikbucks} study correlation clustering with a cheap weak similarity signal and a budget of expensive strong queries.
Cohen-Addad et al.~\cite{cohenaddad2026comparison} ask how many comparisons of two feasible solutions an optimization problem needs when the oracle is exact.

The weak-strong oracle model belongs to a longer line of work on clustering with limited or unreliable supervision.
Ashtiani, Kushagra and Ben-David~\cite{ashtiani2016clustering} introduced same-cluster queries, and Mazumdar and Saha~\cite{mazumdar2017clustering,mazumdar2017side} studied them with persistent noise and proved information-theoretic lower bounds on the number of queries.
Later work made the query bounds tight and the algorithms efficient~\cite{peng2021faulty,xia2022bandit}, and allowed an adversary to flip answers beyond the random noise~\cite{delpia2022semirandom}.
In all of these the noise is on the labels, and the answer to a query is a bit.
In the weak-strong model the noise is on the distances, every pair carries a value, and the strong queries reveal exact distances rather than labels.
Persistent noise, where a repeated query returns the same wrong answer, was studied for sorting by Braverman and Mossel~\cite{braverman2008noisysorting}.

\medskip
MSR was introduced by Doddi et al.~\cite{doddi2000approximation}, and the polynomial-time approximation factors in general metrics are discussed above~\cite{charikar2004clustering,friggstad2022msr,buchem2024msr}.
Ahmadian and Swamy~\cite{ahmadian2016lowerbounds} handle lower bounds on cluster sizes and outliers, and a dynamic version is studied in~\cite{henzinger2020dynamic}.
Parameterizing by $k$ has been effective for MSR and its variants.
Inamdar and Varadarajan~\cite{inamdar2020capacitated} gave the first FPT constant factor for the capacitated version, improved by~\cite{bandyapadhyay2023fpt,jaiswal2024fpt,filtser2026capacitated}.
FPT $(2+\epsilon)$-approximations for MSR were obtained by Chen et al.~\cite{chen2024parameterized}, FPT approximation schemes by Grandoni, Gupta and Yadav~\cite{grandoni2026fptas}, and bounds for fair and other constrained variants in~\cite{banerjee2025improved,bandyapadhyay2025mergeable,nezhad2025fair,gadekar2026parameterized}.
All of these assume that the metric is given exactly.

\subsection{Organization}

\Cref{sec:prelim} defines the problem and the model.
\Cref{sec:overview} gives the algorithm.
\Cref{sec:sampling,sec:rounding,sec:repair,sec:compression} present and analyse the four steps of a test of a guess $\lambda$, and \Cref{sec:guessing} gives the binary search over $\lambda$ and proves \Cref{thm:intromain}.
\Cref{sec:lowerbound} proves the lower bound.

\section{Preliminaries}\label{sec:prelim}
\subsection{Minimum Sum of Radii}
In the \emph{Minimum Sum of Radii} problem (MSR), one is given a set $X$ of $n$ points in a metric space with distances given by $d$, and a positive integer $k$.
We are to pick at most $k$ centres and a radius for each of them, so that every point of $X$ lies within the radius of one of the chosen centres.
Let $n:=|X|$. As in~\cite{bateni2024metric}, distinct points of $X$ have positive distance.
Then, a \emph{ball} in $X$ is a set defined as $B(i,r) = \{j \in X : d(i,j) \leq r\}$ for some centre $i \in X$ and a radius $r \geq 0$.
A solution to the input $X$ is a collection $\mathcal{B}$ of at most $k$ balls $B(i,r)$, $i \in X, r\geq 0$, together with an assignment of every point of $X$ to a ball of $\mathcal{B}$ that contains it.
That is, the algorithm returns the balls and it also says which ball every point belongs to, as in \cite{bateni2024metric}.
The cost of the solution $\mathcal{B}$ is defined as $\mathrm{cost}(\mathcal{B}) = \sum_{B(i,r) \in \mathcal{B}} r$, the total radii of all balls.
We then let $\OPT$ be the minimum of $\mathrm{cost}(\mathcal{B})$ over all possible solutions $\mathcal{B}$.

Fix, only for the analysis, one optimum solution.
That is, $X$ is partitioned into at most $k$ sets $X^*_1,\dots,X^*_k$ with $X^*_i \subseteq B(o_i,r_i)$ and $\sum_i r_i = \OPT$.
We note that the algorithm never sees the sets $X^*_i$, the centres $o_i$, or the radii $r_i$.

\subsection{Weak-Strong Oracle Models}\label{sec:model}
Following Bateni et al.~\cite{bateni2024metric}, we write $\mathsf{Corrupt}\subseteq\binom X2$ for the set of corrupted pairs.
The values of $\Tilde d$ on $\mathsf{Corrupt}$ are chosen by an adversary who knows the metric $(X,d)$ and the set $\mathsf{Corrupt}$.

\subsection{Notations}
We say that an event holds \emph{with high probability} if its probability is at least $1-n^{-\Omega(1)}$.
We write $c$ for a sufficiently large absolute constant, fixed in the analysis, and let $h:=32\lceil c\log n/32\rceil$.
Every witness set in the paper has $h$ points.

The main text uses $\delta=\frac13$.
For general fixed $\delta<\frac12$, \Cref{app:delta} uses $h=\Theta(\log n/(\frac12-\delta)^2)$ and increases the small-component threshold by a factor $1/(\frac12-\delta)$.
The distance tests and radius inflation factors remain unchanged.

The \emph{aspect ratio} of the metric space $(X,d)$ is the largest distance between distinct points divided by the smallest distance between distinct points.
As in~\cite{bateni2024metric}, we assume that the algorithm knows an upper bound $\Delta\le n^{O(1)}$ on the aspect ratio.
This bound does not apply to $\Tilde{d}$ as the adversary could output any value on a corrupted edge.

\section{The algorithm}\label{sec:overview}

Groups of $h=\Theta(\log n)$ strongly queried points serve as witnesses. 
For a witness set $W_s$ around a centre $s$ formed from $h$ closest points to $s$, write $r_s:=\max_{w\in W_s}d(s,w)$ as its radius. 
For $x\notin W_s$, if a strict majority of its weak edges to $W_s$ are uncorrupted, the median weak distance is within $r_s$ of $d(x,s)$. With high probability, for every $h$-point witness set $W$, fewer than $h/2$ points of $X\setminus W$ are corrupt for $W$ (\Cref{lem:fewcorrupt}). 

The algorithm binary searches over a positive guess $\lambda$ of $\OPT/k$ (\Cref{sec:guessing,alg:guessing}). For each guess it runs a \emph{test}, consisting of the following four steps.
For $\OPT/(2k)\le\lambda\le\OPT/k$, the stated outcomes hold with high probability. A point enters the test's local strong set $S$ only when that test has made its query; previously cached exact distances do not change the witness sets or candidate definitions. 

\begin{enumerate}[leftmargin=*]
\item \emph{Sampling (\Cref{sec:sampling}):}
This step identifies candidate balls to be used in forming the final solution. We use an LP whose variables $y_x$ correspond to points $x 
\in X\setminus S$, not yet strongly queried, and whose constraints correspond to candidate balls we have discovered so far; initially we have not discovered any except for trivial $0$-radius balls.

    In a series of rounds, we sample some points based on their LP value (weights $y_x$), strongly query these samples, and use median estimates to generate candidate balls with LP weight greater than their cost to add to the constraints of the LP. One can view this as a sort of probabilistic separation oracle: if the LP weight in a ball of the metric space is too high then a constraint will likely be generated to tamp down the LP solution within that ball in subsequent iterations.

    In total over the rounds, we end up strongly querying $O(k \cdot \log n)$ points and have added sufficiently many candidate balls to reduce the LP value to $O(\OPT)$ (in the case $\lambda \approx\OPT/k$). 
We denote the family of stored candidates by $\mathcal C$, initially $\mathcal C=\varnothing$. The trivial singleton choices are represented by the individual LP caps.

\item \emph{Rounding (\Cref{sec:rounding}):}
This step uses a dual-growing process on the final LP (in the last iteration of sampling including all the candidate balls it accepted as its constraints) to create a shorter list of candidate balls and a small number of singleton balls whose total cost (paying radius plus $\lambda$ for each selected candidate ball, and $\lambda/h$ for each singleton ball) is $O(\OPT)$. 
After strongly querying the point in each singleton ball, the other selected balls are then merged into $O(k)$ candidate ``components'' each of which is covered by a single ball. Every point that has not been strongly queried is assigned to be covered by the ball around some candidate component.
The {\em list} of a component is the set of unqueried points assigned to it.

However, these new assignments must yet again be verified to maintain the high probability property.

\item \emph{Repair (\Cref{sec:repair}):}
This step corrects the assignments that rounding may have got wrong, call it set $E$ over the test. 
We strongly query points with conflicting estimates and use the global bound $|E|=O(kh)$ to limit their number. We strongly query every component whose list is small compared with a bound on the number of points in the sets $E_s$ of its own centres and on the number of points already queried. For each larger component, we test points using $h$ fresh witnesses, possibly reassigning points to new components and forming new lists. A uniform bound over unions of true balls justifies these new witnesses despite the earlier weak-dependent assignments. With high probability, the points in each new list are covered by the ball for that component. Repair uses $O(k\log n)$ strong queries in total.

\item \emph{Compression (\Cref{sec:compression}):}
At this point we have with high probability a solution covering the points not yet strongly queried using $O(k)$ balls of total cost $O(\OPT)$. This step merges them into just $k$ balls of total cost $O(\OPT)$ by running a constant-factor approximation on the strongly queried points and using them to merge the $O(k)$ balls into just $k$ balls.

Each list is assigned through its strongly queried root to a ball of the offline solution. We enlarge that ball using the proved covering radius of the component; this does not require querying distances to the unqueried members of the list.

\end{enumerate}

Each detailed procedure checks its query and cardinality budgets before making queries. Thus every test, including one that aborts, uses $O(kh)$ strong queries. A test passes only if the returned balls have total radius at most $Ck\lambda$, with $C=O(1)$ as in \Cref{lem:compression}. On the simultaneous event in the proof of \Cref{lem:binsearch}, every passing test is valid, and a test with $\lambda\ge\OPT/(2k)$ does not abort. The search uses $O(\log\log n)$ tests, giving an $O(1)$-approximation with $O(k\log n\log\log n)$ strong queries.

\subsection{Sampling}\label{sec:sampling}

The first step constructs a family of \emph{candidates}, a set of centre-radius pairs, from which we build the final solution. 
With high probability, removing fewer than $h/2$ points from each candidate leaves a subset of its associated true metric ball (\Cref{lem:sampling}). A packing linear program assigns weight to points that the current candidates do not adequately account for. Sampling according to this weight identifies new centres; \Cref{lem:sampling} shows that the final LP objective is small with high probability. This step uses $O(k\log n)$ strong queries.

Fix a positive guess $\lambda$. Throughout this subsection, $S$
denotes the points already strongly queried; exact distances are
therefore known between every pair of points of $S$. We use $h$
witnesses per sampled centre, where $h$ is an even integer with
$h=\Theta(\log n)$, and choose the absolute constant $c$ in the
algorithm sufficiently large. 

A candidate is a triple $(A(s,R),s,R)$, with $s\in S$ and
$A(s,R)\subseteq X\setminus S$. 
Its set $A(s,R)$ is computed from
weak-distance estimates, whereas $B(s,R)$ always denotes a ball
in the true metric. We sometimes refer to a candidate by its set
when the centre and radius are understood.
Note that $A(s,R)$ and $B(s,R)$ differ only in corrupted points for $W_s$, called $E_s$. 

The main challenge in sampling is deciding where to make the
next strong queries. We would like to focus on regions that are
not yet adequately represented by the candidates already found.
Uniform sampling does not achieve this: a region that is important
for an optimal solution may contain relatively few points, while
many points may lie in regions already represented by candidates.
We therefore use a linear program to assign weights to the
remaining points, with the goal of directing sampling toward
regions that still need attention.

For a candidate family $\mathcal C$, consider the packing LP
\begin{equation}\label{eq:packing}
\begin{aligned}
\max\quad &\sum_{x\in X\setminus S}y_x\\
\text{subject to}\quad
&y(A(s,R))\le R+\lambda
&&((A(s,R),s,R)\in\mathcal C)&&\text{(candidate constraints)},\\
&0\le y_x\le\frac{\lambda}{h}
&&(x\in X\setminus S)&&\text{(cap constraints)},
\end{aligned}
\end{equation}
where $y(A)=\sum_{x\in A}y_x$. 

To understand this LP, it is helpful to view it as the dual of a fractional covering problem. For a point $i\in X$ and a radius $r$, the variable $z_{i,r}$ indicates that we pick the ball $B(i,r)$.
Below, the first LP is the standard LP relaxation of MSR.
The second moves the constraint $\sum_{i,r}z_{i,r}\le k$ into the objective with multiplier $\lambda$, and for every $\lambda\ge0$ it gives a lower bound on the first.
The third is the dual of the second.

\begin{center}
\begin{tabular}{@{}c@{\hspace{1.2em}}c@{\hspace{1.2em}}c@{}}
$\begin{aligned}[t]
\min\ &\textstyle\sum_{i,r}r\,z_{i,r}\\
\text{s.t.}\ &\textstyle\sum_{i,r:\,x\in B(i,r)}z_{i,r}\ge1\ \ \forall x\\
&\textstyle\sum_{i,r}z_{i,r}\le k\\
&z\ge0
\end{aligned}$
&
$\begin{aligned}[t]
\min\ &\textstyle\sum_{i,r}(r+\lambda)z_{i,r}-\lambda k\\
\text{s.t.}\ &\textstyle\sum_{i,r:\,x\in B(i,r)}z_{i,r}\ge1\ \ \forall x\\
&z\ge0
\end{aligned}$
&
$\begin{aligned}[t]
\max\ &\textstyle\sum_{x\in X}y_x-\lambda k\\
\text{s.t.}\ &y(B(i,r))\le r+\lambda\ \ \forall i,r\\
&y\ge0
\end{aligned}$
\\[1ex]
\emph{Primal LP} & \emph{Lagrangian LP} & \emph{Dual LP}
\end{tabular}
\end{center}

In that problem, we may pay
$R+\lambda$ to select a candidate of radius $R$, or pay a
penalty of $\lambda/h$ for an individual point that is not
covered by the selected candidates. The variables $y_x$
can therefore be interpreted as weights placed on points:
the total weight inside a candidate cannot exceed its cost
$R+\lambda$, and no individual point can carry more than
$\lambda/h$. The parameter $\lambda$ acts as an opening cost
for each candidate, in addition to its radius. This makes
$\OPT+k\lambda$ the natural benchmark, corresponding to the
total radius of an optimal solution plus an opening cost
for each of its at most $k$ balls.
Without the cap $\lambda/h$, and with a constraint for every ball, this LP is the dual of the Lagrangian relaxation of the standard LP for MSR.
The cap ensures that a cluster with large weight has many points, so draws into it reach many distinct points.

LP~\eqref{eq:packing} is the dual LP without the constant $-\lambda k$ and with the cap constraints added.
It has variables only for the points of $X\setminus S$, and candidate constraints only for the candidates in $\mathcal C$.
At the start $\mathcal C$ is empty, and it grows over the rounds of sampling.

We write $\OPTLP$ for the optimum of this LP at the
\emph{end} of sampling. The following summarizes the guarantees we will prove in this section.

\begin{lemma}\label{lem:sampling}
The sampling step makes $O(k\log n)$ strong queries and returns a candidate family $\mathcal C$ such that, with high probability,
\begin{enumerate}
\item for each $(A(s,R),s,R)\in\mathcal C$, some set $E_s$ with $|E_s|<h/2$ satisfies $A(s,R)\setminus E_s\subseteq B(s,R)$; and
\item $\OPTLP<17(\OPT+k\lambda)$.
\end{enumerate}
\end{lemma}

\subsubsection{The sampling procedure}

Initially $S=\varnothing$ and $\mathcal C=\varnothing$. We perform at most $ck$ rounds, stopping earlier if $|X\setminus S|\le ckh$; in that case we strongly query all remaining points. In a round, solve \eqref{eq:packing}, independently draw $h$ points with replacement with probabilities proportional to an optimal solution $y$, and let $Q$ be the set of distinct drawn points. Strongly query $Q$ and add it to $S$.

\begin{figure}[ht]
\centering
\begin{tikzpicture}[scale=1.15,
   pt/.style={circle,fill=black!75,inner sep=2pt},
   ctr/.style={circle,fill=blue,inner sep=2.4pt},
   wit/.style={circle,fill=blue,inner sep=1.3pt},
   lbl/.style={font=\small}]
  \draw[fill=blue!6,draw=blue!45,dashed] (0,0) circle (1.4);
  \node[ctr] (s) at (0,0) {};
  \node[lbl,text=blue] at (0,-0.32) {$s$};
  \foreach \a/\r in {350/1.15, 75/0.9, 130/1.2, 200/0.95, 258/1.25, 320/0.8}{
     \node[wit] (w\a) at (\a:\r) {};
  }
  \draw[blue,{Latex[length=1.5mm]}-{Latex[length=1.5mm]}] (0,0) -- (105:1.4);
  \node[lbl,text=blue] at (-0.38,0.70) {$r_s$};
  \node[pt] (x) at (4.0,0.9) {};
  \node[lbl] at (4.26,1.12) {$x$};
  \foreach \a in {350,258,320,75}{
     \draw[black!45,thin,dashed,-{Latex[length=1.3mm]}] (x) -- (w\a);
  }
  \draw[black!80] (s) -- node[lbl,sloped,pos=0.55,fill=white,inner sep=1pt] {$d(s,x)$} (x);
\end{tikzpicture}
\caption{The witness set and the estimated distance.
The dashed edges are weak edges, and some of them may be corrupted.
The witness set $W_s$ (blue) holds the $h$ nearest points of $S\setminus\{s\}$ to $s$, all within distance $r_s$.
For $x\notin S\cup E_s$, the median of its weak distances to $W_s$, shifted by $r_s$, lies between $d(s,x)$ and $d(s,x)+2r_s$ (\Cref{lem:proxy}).}
\label{fig:witness}
\end{figure}
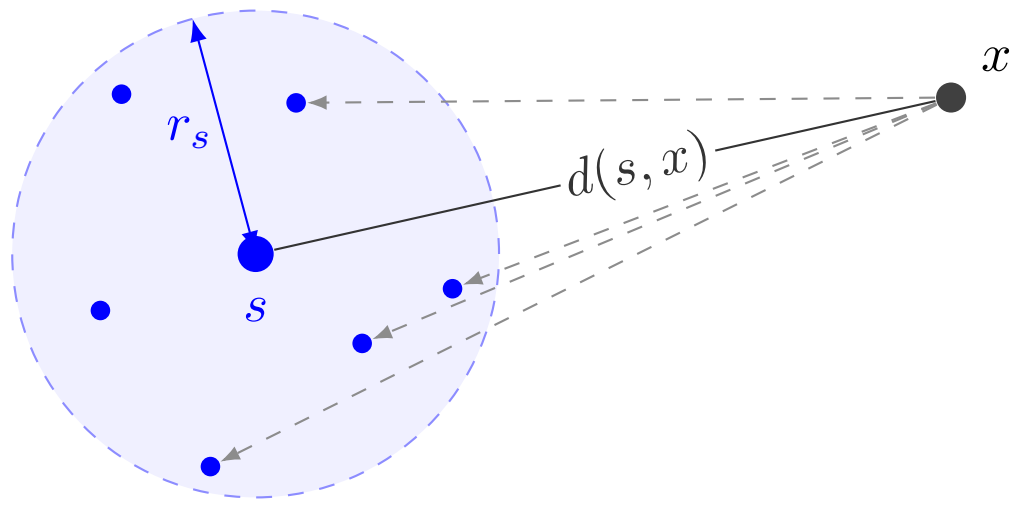

When $|S|\ge h+1$, each $s\in Q$ obtains a witness set $W_s$ consisting of its $h$ nearest neighbours in $S\setminus\{s\}$, measured using exact distances. Define
\begin{equation}\label{eq:witnessradius}
 r_s:=\max_{w\in W_s}d(s,w).
\end{equation}
For $x\notin S$ define the estimated distance
\begin{equation}\label{eq:score}
 \md(x,s):=\max\!\left\{0,\operatorname{median}_{w\in W_s}\Tilde d(x,w)\right\}+r_s,
\end{equation}
and set $\md(x,s)=d(x,s)$ for $x\in S$. The median suppresses a minority of corrupted edges; adding $r_s$ makes the estimate an upper bound on the true distance whenever the majority is honest (see \Cref{fig:witness}).

Let $S_{\rm old}:=S\setminus Q$ denote the strong set at the
start of the round.

For each newly queried point $s\in Q$ for which a witness set has been formed, and each radius $R\ge0$,
temporarily define $A(s,R):=\{x\in X\setminus S_{\rm old}:\md(x,s)\le R\}$.

Although $R$ can take any nonnegative value, we need
to examine only finitely many radii for each fixed centre
$s$. Indeed, as $R$ increases, the set $A(s,R)$ changes
only when $R$ reaches the estimated distance $\md(x,s)$
of some point $x\in X\setminus S_{\rm old}$. Thus it suffices to consider $\mathcal R_s
 :=\{\md(x,s):x\in X\setminus S_{\rm old}\}$, which contains at most $n$ distinct values. Since $|Q|\le h$, at most
$nh$ centre-radius pairs are examined in each round.

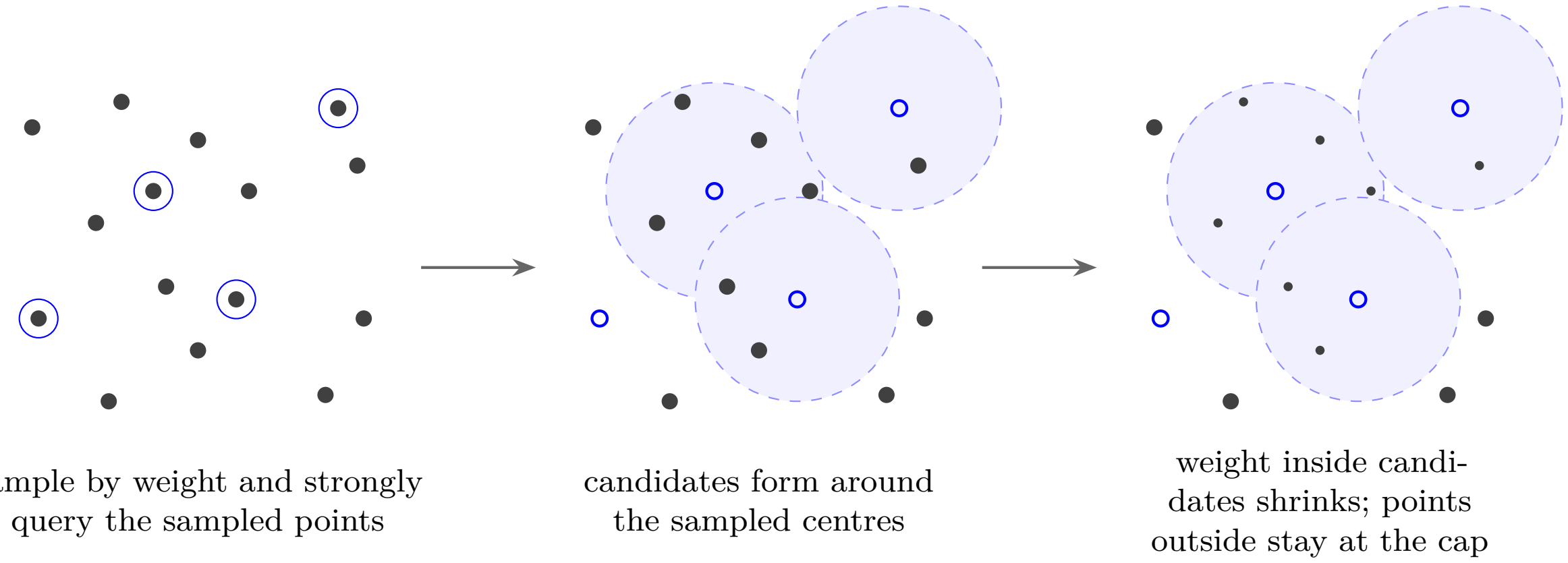
\begin{figure}[ht]
\centering
\begin{tikzpicture}[scale=1.25,
   pt/.style={circle,fill=black!75,inner sep=1.6pt},
   ptS/.style={circle,fill=black!75,inner sep=0.9pt},
   ctr/.style={circle,draw=blue,thick,inner sep=1.5pt},
   hit/.style={circle,draw=blue,inner sep=3.8pt},
   cand/.style={draw=blue!45,dashed,fill=blue!6},
   capt/.style={font=\footnotesize,text width=4.4cm,align=center}]

\begin{scope}
  \foreach \x/\y in {0.3/2.2, 1.0/2.4, 1.6/2.1, 0.8/1.45, 2.0/1.7, 2.85/1.9,
                     1.35/0.95, 2.9/0.7, 1.6/0.45, 2.6/0.1, 0.9/0.05}{
     \node[pt] at (\x,\y) {};
  }
  \foreach \x/\y in {1.25/1.7, 1.9/0.85, 2.7/2.35, 0.35/0.7}{
     \node[pt] at (\x,\y) {};
     \node[hit] at (\x,\y) {};
  }
  \node[capt] at (1.6,-0.75) {sample by weight and strongly query the sampled points};
\end{scope}

\draw[-{Stealth},thick,black!60] (3.35,1.1) -- (4.25,1.1);

\begin{scope}[shift={(4.4,0)}]
  \draw[cand] (1.25,1.7) circle (0.85);
  \draw[cand] (1.9,0.85) circle (0.8);
  \draw[cand] (2.7,2.35) circle (0.8);
  \foreach \x/\y in {0.3/2.2, 1.0/2.4, 1.6/2.1, 0.8/1.45, 2.0/1.7, 2.85/1.9,
                     1.35/0.95, 2.9/0.7, 1.6/0.45, 2.6/0.1, 0.9/0.05}{
     \node[pt] at (\x,\y) {};
  }
  \foreach \x/\y in {1.25/1.7, 1.9/0.85, 2.7/2.35, 0.35/0.7}{
     \node[ctr] at (\x,\y) {};
  }
  \node[capt] at (1.6,-0.75) {candidates form around the sampled centres};
\end{scope}

\draw[-{Stealth},thick,black!60] (7.75,1.1) -- (8.65,1.1);

\begin{scope}[shift={(8.8,0)}]
  \draw[cand] (1.25,1.7) circle (0.85);
  \draw[cand] (1.9,0.85) circle (0.8);
  \draw[cand] (2.7,2.35) circle (0.8);
  \foreach \x/\y in {1.0/2.4, 1.6/2.1, 0.8/1.45, 2.0/1.7, 2.85/1.9, 1.35/0.95, 1.6/0.45}{
     \node[ptS] at (\x,\y) {};
  }
  \foreach \x/\y in {0.3/2.2, 2.9/0.7, 2.6/0.1, 0.9/0.05}{
     \node[pt] at (\x,\y) {};
  }
  \foreach \x/\y in {1.25/1.7, 1.9/0.85, 2.7/2.35, 0.35/0.7}{
     \node[ctr] at (\x,\y) {};
  }
  \node[capt] at (1.6,-0.75) {weight inside candidates shrinks; points outside stay at the cap};
\end{scope}

\end{tikzpicture}
\caption{One sampling round.
The size of a point is its LP weight.
Left to right: the sampled points (circled), the candidates around them, and the new LP weights.}
\label{fig:samplinground}
\end{figure}

For each such value $R$, add $(A(s,R),s,R)$ to $\mathcal C$ if
\begin{equation}\label{eq:accept}
 y(A(s,R))>R+\lambda.
\end{equation}
Here $y$ is the solution \emph{before} the new queries, so the temporary sets include the points of $Q$ and retain their weights. At the end of the round, intersect the set of every stored candidate with the new $X\setminus S$; its centre and radius are unchanged. Consequently, every candidate constraint refers to precisely the points that remain unqueried. The outer loop is specified above, and one round is given in \Cref{alg:sampling}. \Cref{fig:samplinground} illustrates one round.

\begin{algorithm}[H]
\caption{\textup{\textsc{Sample}}$(S,\mathcal C,\lambda)$: one sampling round.}\label{alg:sampling}
\LinesNumbered
\KwIn{The current strong set $S$, candidate family $\mathcal C$, and guess $\lambda$}
Solve \eqref{eq:packing} to obtain $y$; let $S_{\rm old}\gets S$\;
Draw $h$ independent samples proportional to $y$; let $Q$ be their set of distinct points\;
Strongly query the points of $Q$ and insert them into $S$ in their order of first appearance among the draws\;
\If{$|S|\ge h+1$}{
 \For{each $s\in Q$}{
  $W_s\gets$ the $h$ nearest points to $s$ in $S\setminus\{s\}$, breaking ties by insertion order in $S$; compute $r_s$ and $\md(\cdot,s)$\;
  \For{each distinct $R\in\{\md(x,s):x\in X\setminus S_{\rm old}\}$}{
   $A\gets\{x\in X\setminus S_{\rm old}:\md(x,s)\le R\}$\;
   \lIf{$y(A)>R+\lambda$}{add $(A,s,R)$ to $\mathcal C$}
  }
 }
}
Replace every stored $(A,s,R)$ by $(A\cap(X\setminus S),s,R)$\;
\KwRet{$S,\mathcal C$}\;
\end{algorithm}

\subsubsection{Uniform control of corrupted estimates}

Because witness sets are chosen adaptively, concentration for a \emph{fixed} witness set is not enough. Instead, we establish a simultaneous guarantee for every possible witness set.

\begin{definition}\label{def:corrupt}
Let $W\subseteq X$ have size $h$.
A point $x\notin W$ is \emph{corrupt for $W$} if at least $h/2$ of its edges to $W$ are in $\mathsf{Corrupt}$.
For a sampled centre $s$ with witness set $W_s$, let $E_s$ be the set of points corrupt for $W_s$.
We call a point \emph{corrupt} if it is corrupt for some $W_s$, and we let $E:=\bigcup_sE_s$ be the set of corrupt points.
\end{definition}

\begin{lemma}\label{lem:fewcorrupt}
For a sufficiently large constant $c$, with high probability, every set $W\subseteq X$ of size $h$ has fewer than $h/2$ points corrupt for $W$.
\end{lemma}
\begin{proof}
Fix disjoint $W,M\subseteq X$ of size $h$ and $h/2$, respectively. For any $x\in M$, the $h$ edges from $x$ to $W$ are corrupted independently, each with probability at most $1/3$. \Cref{lem:hoeffding} gives
\[
 \Pr[x\text{ is corrupt for }W]\le \exp(-h/18).
\]
The edge sets used by distinct $x\in M$ are disjoint, so
\[
 \Pr[\text{all points of }M\text{ are corrupt for }W]\le\exp(-h^2/36).
\]
There are at most $n^{3h/2}\le n^{2h}$ choices of $(W,M)$. Hence the probability that any such pair exists is at most
\[
 n^{2h}\exp(-h^2/36)=n^{-\Omega(1)}
\]
when $h\ge c\log n$ and $c$ is sufficiently large. If at least $h/2$ points were corrupt for a witness set, one of these pairs would exist.
\end{proof}

The event of \Cref{lem:fewcorrupt} depends only on the set $\mathsf{Corrupt}$, not on the adversarial values assigned to corrupted edges, and it holds simultaneously for adaptively chosen witness sets. On this event, $|E_s|<h/2$, independently of later changes to $S$.

The individual bounds do not by themselves give the sharp bound on their union. We also use the fact that sampling makes only $O(kh)$ strong queries.

\begin{lemma}\label{lem:globalcorrupt}
For a sufficiently large constant $c$, with high probability $|E|\le2ckh$ in one execution.
\end{lemma}
\begin{proof}
If sampling queries all leftover points, then $|X|\le2ckh$ and the claim is immediate.
Otherwise sampling runs all $ck$ rounds and makes exactly $ckh$ draws.
Let $S$ be the strong set at the end of sampling, so $|S|\le ckh$.

Fix a sequence of $ckh$ draws, and fix a set $M\subseteq X\setminus S$ with $|M|=ckh$.
Each round makes $h$ draws, so the sequence and the true metric determine $S$ and all of the at most $ckh$ witness sets.
For $x\in M$ and a witness set $W_s$, \Cref{lem:hoeffding} gives
\[
 \Pr[x\text{ is corrupt for }W_s]\le e^{-h/18}.
\]
A union bound over the witness sets gives
\[
 \Pr[x\in E]\le ckh\,e^{-h/18}.
\]
Distinct points of $M$ use disjoint edge sets into $S$, so
\[
 \Pr[M\subseteq E]\le(ckh\,e^{-h/18})^{ckh}.
\]
There are at most $n^{ckh}$ sequences and at most $n^{ckh}$ sets $M$, and $ckh\le n$ because $M$ exists.
So the probability that some sequence and some set $M$ have this property is at most
\[
 n^{2ckh}(ckh\,e^{-h/18})^{ckh}\le n^{3ckh}e^{-ckh^2/18}=n^{-\Omega(1)},
\]
when $h\ge c\log n$ and $c$ is sufficiently large.
The bound covers the actual sequence of draws, even though it depends on the weak distances.
So with high probability fewer than $ckh$ points outside $S$ lie in $E$.
Adding the at most $ckh$ points of $S$ proves the claim.
\end{proof}

We need the next lemma because candidates are built from estimated distances, not true distances.
It shows that every candidate $A(s,R)$ lies inside the true ball $B(s,R)$, except for the fewer than $h/2$ points of $E_s$.

\begin{lemma}\label{lem:proxy}
For every sampled centre $s$ with a witness set and every $x\notin E_s$,
\begin{equation}\label{eq:proxy}
 d(s,x)\le\md(x,s)\le d(s,x)+2r_s.
\end{equation}
Consequently $A(s,R)\setminus E_s\subseteq B(s,R)$ for every candidate at $s$. On the event of \Cref{lem:fewcorrupt}, we also have $|E_s|<h/2$.
\end{lemma}
\begin{proof}
The assertion is immediate for a strongly queried point $x$, whose distance estimate is exact. Otherwise, strictly more than half the edges from $x$ to $W_s$ are honest. Each honest witness $w$ satisfies
\[
 d(s,x)-r_s\le d(x,w)\le d(s,x)+r_s.
\]
The median $M$ of the weak distances therefore lies in the same interval. If $d(s,x)\ge r_s$, then $M\ge0$ and $M+r_s\ge d(s,x)$; if $d(s,x)<r_s$, then $\max\{0,M\}+r_s\ge r_s>d(s,x)$. In both cases,
\[
 d(s,x)\le\max\{0,M\}+r_s\le d(s,x)+2r_s.
\]
The remaining claims follow from \Cref{lem:fewcorrupt} and the definition of $A(s,R)$.
\end{proof}

This establishes the first assertion of \Cref{lem:sampling} on the event of \Cref{lem:fewcorrupt}. On the same event, the LP caps give
\begin{equation}\label{eq:badweight}
 y(E_s\cap(X\setminus S))\le |E_s|\frac{\lambda}{h}<\frac{\lambda}{2}.
\end{equation}
We write $y(E_s)$ for the left-hand side when the LP domain is clear.

\subsubsection{Heavy clusters and captured clusters}

Fix an optimal solution with at most $k$ clusters $X_i^*\subseteq B(o_i,r_i)$ and $\sum_i r_i=\OPT$. At the start of a round, put $F_i:=X_i^*\setminus S$, and let $y$ be the LP solution used in that round.

\begin{definition}\label{def:heavy}
For the constant $\alpha=8$, cluster $i$ is \emph{heavy} if
\begin{equation}\label{eq:heavy}
 y(F_i)>\alpha(r_i+\lambda).
\end{equation}
\end{definition}

\begin{remark}\label{rem:light}
The total weight of nonheavy clusters is at most
\begin{equation}\label{eq:light}
 \sum_{i\text{ nonheavy}}y(F_i)\le\alpha\sum_i(r_i+\lambda)
 \le\alpha(\OPT+k\lambda).
\end{equation}
\end{remark}

\begin{definition}\label{def:captured}
Cluster $i$ is \emph{captured} if the current family contains a candidate $(A(s,R),s,R)$ with $R\le6r_i$ and $F_i\setminus E_s\subseteq A(s,R)$.
\end{definition}

This property persists as sampling proceeds: both $F_i$ and the set of points in a stored candidate lose any newly queried points. On the event of \Cref{lem:fewcorrupt}, capture prevents a cluster from becoming heavy again (\Cref{lem:capturednotheavy}).

\begin{lemma}\label{lem:capturednotheavy}
On the event of \Cref{lem:fewcorrupt}, every captured cluster is nonheavy for every feasible solution of the current packing LP.
\end{lemma}
\begin{proof}
If $(A(s,R),s,R)$ captures cluster $i$, the candidate constraint and \eqref{eq:badweight} imply
\[
 y(F_i)\le y(A(s,R))+y(E_s)\le R+\frac{3\lambda}{2}
 \le6r_i+\frac{3\lambda}{2}\le8(r_i+\lambda).
\]
Thus the strict inequality in \eqref{eq:heavy} cannot hold.
\end{proof}

\subsubsection{Why repeated sampling makes progress}

Repeated draws from a heavy cluster give sufficiently many distinct points with high probability (\Cref{lem:distinct,lem:hits}). On the event of \Cref{lem:fewcorrupt}, a round that samples the cluster after sufficiently many distinct points have been queried constructs a candidate that captures it (\Cref{lem:capture}).

\begin{lemma}\label{lem:capture}
On the event of \Cref{lem:fewcorrupt}, consider a round and suppose that:
\begin{enumerate}
\item cluster $i$ is heavy at the start of the round;
\item the round draws a point $s\in F_i$; and
\item after all draws, at least $h$ points of $X_i^*$ lie in $S\setminus\{s\}$.
\end{enumerate}
Then this round adds to $\mathcal C$ a candidate capturing cluster $i$.
\end{lemma}
\begin{proof}
Every point of $X_i^*$ is within $2r_i$ of $s$. Since $S\setminus\{s\}$ contains at least $h$ such points, the nearest-witness radius satisfies $r_s\le2r_i$. By \Cref{lem:proxy},
\[
 \md(x,s)\le d(s,x)+2r_s\le6r_i
 \qquad(x\in F_i\setminus E_s).
\]
Choose $R$ to be the largest of these estimates, so $R\le6r_i$ and $F_i\setminus E_s\subseteq A(s,R)$. This maximum exists: heaviness and the cap imply that $F_i\setminus E_s$ has positive LP weight. The round considers this radius. Moreover, the temporary candidate includes the points drawn in this round, so the LP solution from the start of the round gives
\[
 y(A(s,R))\ge y(F_i)-y(E_s)
 >8(r_i+\lambda)-\frac\lambda2>R+\lambda.
\]
Thus \eqref{eq:accept} holds. After sampled points are deleted from the candidate, it still captures the current $F_i$.
\end{proof}

The next sampling bound quantifies the number of distinct points obtained despite replacement.

\begin{lemma}\label{lem:distinct}
Let $p$ be a probability distribution with $\max_x p_x\le1/(8m)$. In $2m$ independent draws from $p$,
\[
 \Pr[\text{at most $m$ distinct outcomes}]\le2^{-m}.
\]
The same conclusion holds for sequential draws whose conditional probability of each outcome is always at most $1/(8m)$.
\end{lemma}
\begin{proof}
A draw is a repeat if its outcome has appeared earlier. On a sequence containing at most $m$ distinct outcomes, at least $m$ of the $2m$ draws are repeats. As long as at most $m$ values have appeared, the next draw repeats one of them with conditional probability at most $m/(8m)=1/8$. For any prescribed $m$ repeat positions, the probability that all are repeats (while at most $m$ distinct outcomes have appeared) is at most $8^{-m}$. A union bound over the at most $\binom{2m}{m}\le4^m$ choices of positions yields $2^{-m}$. This argument only uses conditional probability bounds, giving the sequential version as well.
\end{proof}

\begin{lemma}\label{lem:hits}
Condition on the event of \Cref{lem:fewcorrupt}. With probability at least $1-k2^{-h}$ over the sampling randomness, the following holds for every optimal cluster $i$: consider all rounds at whose start cluster $i$ is heavy. Cluster $i$ receives at most $3h$ draws in total over these rounds.
\end{lemma}
\begin{proof}
Fix $i$ and examine, in chronological order, draws landing in $F_i$ during rounds in which $i$ is heavy. Conditioned on such a draw occurring, its probability of selecting a particular $x\in F_i$ is
\[
 \frac{y_x}{y(F_i)}\le\frac{\lambda/h}{8(r_i+\lambda)}\le\frac1{8h}.
\]
This inequality continues to hold conditioned on all previous draws, including draws in earlier rounds. By the sequential form of \Cref{lem:distinct}, if the first $2h$ heavy-cluster draws occur, they contain at least $h+1$ distinct points except with probability $2^{-h}$. In the round containing the $2h$-th draw, choose that draw's point as $s$. The cluster is heavy at the start of that round, and $S\setminus\{s\}$ contains at least $h$ points of $X_i^*$. By \Cref{lem:capture}, the round captures the cluster. By \Cref{lem:capturednotheavy}, no later round sees it heavy. There were fewer than $2h$ such draws before the capturing round and at most $h$ in that round. Hence their total is at most $3h$. Take a union bound over the at most $k$ clusters.
\end{proof}

In contrast, if the LP objective is large, most of its weight must lie in heavy clusters, so a round makes many such draws.

\begin{lemma}\label{lem:round}
Condition on the history before a round, and suppose its LP optimum is at least $(2\alpha+1)(\OPT+k\lambda)$. If $N$ counts that round's fresh draws into clusters heavy at its start, then, over those draws,
\[
 \Pr[N\ge h/4]\ge1-e^{-h/16}.
\]
\end{lemma}
\begin{proof}
By \eqref{eq:light}, the nonheavy clusters account for at most $\alpha(\OPT+k\lambda)$, strictly less than half the assumed objective. Each of the $h$ independent draws therefore lands in a heavy cluster with probability greater than $1/2$. The lower-tail bound in \Cref{lem:chernoff}, with $\epsilon=1/2$, gives $\Pr[N<h/4]\le e^{-h/16}$.
\end{proof}

\begin{proof}[Proof of \Cref{lem:sampling}]
The first assertion follows from \Cref{lem:proxy}. For the second, fix a corruption pattern satisfying \Cref{lem:fewcorrupt} and any weak-distance table assigned by the adversary to that pattern. All probabilities in the following argument are over the sampling randomness. Observe that the LP optimum never increases: a subsequent round only adds candidate constraints and removes queried variables from every stored set. Extending a feasible solution of the new LP by zero on removed variables gives a feasible solution of the old LP with the same objective.

If sampling stops because at most $ckh$ points remain, those points are all queried and the final LP has optimum $0$. Otherwise all $ck$ rounds occur. Suppose that the final LP optimum is at least $(2\alpha+1)(\OPT+k\lambda)$. By monotonicity, each round started above this threshold. Applying \Cref{lem:round} in every round, except with probability at most $ck e^{-h/16}$ at least $h/4$ draws per round landed in clusters heavy at their respective starts. This would yield at least $ckh/4>3kh$ heavy-cluster draws when $c>12$, contradicting \Cref{lem:hits}, except on an event of probability at most $k2^{-h}$. Hence
\[
 \Pr\!\left[\OPTLP\ge(2\alpha+1)(\OPT+k\lambda)\right]
 \le ck e^{-h/16}+k2^{-h}=n^{-\Omega(1)}.
\]
With $\alpha=8$, we obtain $\OPTLP<17(\OPT+k\lambda)$ except with the conditional failure probability just displayed. Removing the conditioning adds at most the failure probability in \Cref{lem:fewcorrupt}, so both assertions hold with high probability.

Each of the at most $ck$ rounds queries at most $h$ new points, and the optional final step queries at most $ckh$ more. The total is at most $2ckh=O(k\log n)$.
\end{proof}

\subsection{Rounding}\label{sec:rounding}

The sampling procedure may produce many overlapping candidates. For $\lambda\ge\OPT/(2k)$, with high probability rounding selects only $O(k)$ of them, with total radius $O(k\lambda)$, and assigns the remaining unqueried points to components of selected balls (\Cref{lem:rounding}). Since the candidates are defined using weak distances, these assignments may initially be wrong for corrupt points. The next subsection repairs them.

Overall, we prove the following results in this section.
\begin{lemma}\label{lem:rounding}
If $\lambda\ge\OPT/(2k)$, then, with high probability, the rounding step does not abort, makes $O(k\log n)$ additional strong queries, and partitions the points it leaves unqueried into at most $O(k)$ nonempty lists $X_J$. On this event, each list has an associated ball $B_J=B(c_J,R_J)$, where $c_J$ is strongly queried, such that
\begin{enumerate}
\item $\sum_JR_J=O(k\lambda)$; and
\item $X_J\setminus E\subseteq B_J$, where the set $E$ has size $O(kh)=O(k\log n)$.
\end{enumerate}
\end{lemma}

\subsubsection{Growing, selecting, and merging candidates}

During the primal--dual growth phase, $S$ and $\mathcal C$ refer to their values \emph{at the end of sampling}; they remain fixed until this phase finishes. A variable $y_x$ is frozen once it reaches its cap or participates in a tight candidate constraint; otherwise it remains unfrozen. The growth and selection steps follow the primal--dual viewpoint used in offline MSR approximations~\cite{charikar2004clustering,buchem2024msr}; the additional repair step handles the errors in weak-distance candidates.

Let $Z$ be the initially empty family of candidates that become tight during growth. Starting from $y=0$, increase every unfrozen variable at the same rate until either a candidate constraint becomes tight or a variable reaches its cap $\lambda/h$. In the first case, add the candidate to $Z$ and freeze all its points; in the second, freeze the point and put it into a set $L$ of \emph{singletons}. Resolve simultaneous events consistently, processing all newly tight constraints. Every variable eventually freezes, and the resulting $y$ is feasible for \eqref{eq:packing}.

The singleton points will be queried exactly. For the other points, two reductions simplify $Z$. First, \emph{prune} candidates with a common centre, retaining the largest radius. Next, process the pruned candidates in nonincreasing order of radius, and select one only if its centre is at distance greater than the sum of the two radii from every centre selected earlier. Denote this disjoint subfamily by
\[
 Z'=\{(A(s_i,R_i),s_i,R_i):1\le i\le m\}.
\]
Every candidate discarded during selection is assigned to an earlier selected candidate whose centre is at distance at most the sum of their radii from its centre. A candidate discarded in pruning inherits the assignment of the larger candidate at its centre. Because the centres are strongly queried, all tests of distances between centres are exact.
\Cref{fig:rounding} illustrates growth and selection.

Finally, inflate the selected balls to $B(s_i,5R_i)$ and connect $i$ and $j$ when $d(s_i,s_j)\le7(R_i+R_j)$. For each connected component $J$ choose one of its centres as a root $c_J$ and set
\[
 R_J:=14\sum_{i\in J}R_i,\qquad B_J:=B(c_J,R_J).
\]
Assign every point frozen through a candidate to the component reached by its selected representative, and let $X_J$ be the resulting list.
We also let $E_J:=\bigcup_{i\in J}E_{s_i}$ be the set of points that are corrupt for the witness set of some centre in $J$ (\Cref{def:corrupt}).
Strongly queried singletons are removed from the lists and added to $S$. \Cref{fig:merging} illustrates the merging. The complete procedure is in \Cref{alg:rounding}.

\begin{algorithm}[H]
\caption{\textup{\textsc{Round}}: rounding the candidate family.}\label{alg:rounding}
\LinesNumbered
\KwIn{The candidate family $\mathcal C$, current strong set $S$, and guess $\lambda$}
Set $y_x\gets0$ for $x\in X\setminus S$; set $Z,L,Z'\gets\varnothing$\;
\While{some variable is unfrozen}{
 Raise all unfrozen $y_x$ uniformly until a candidate constraint or a cap becomes tight\;
 Freeze all points of each newly tight candidate and add those candidates to $Z$\;
 Freeze any point reaching $\lambda/h$ and add it to $L$ if it is not already frozen through a candidate\;
}
\lIf{$|L|>ckh$}{\textbf{abort}}
Strongly query the points of $L$ (but keep the LP domain fixed)\;
Prune $Z$, keeping only the largest-radius candidate at each centre\;
\For{each remaining $(A(s,R),s,R) \in Z$, in nonincreasing order of $R$}{
 \lIf{some $(A(s',R'),s',R')\in Z'$ satisfies $d(s,s')\le R+R'$}{assign the candidate to such a selected candidate}
 \lElse{insert the candidate into $Z'$}
}
\lIf{$|Z'|>ck$}{\textbf{abort}}
Build the graph on $Z'$ with edges $ij$ when $d(s_i,s_j)\le7(R_i+R_j)$\;
\For{each connected component $J$}{
 Choose any one $c_J\in\{s_i:i\in J\}$ and put $R_J\gets14\sum_{i\in J}R_i$\;
}
Assign every point of $X\setminus(S\cup L)$ through its selected representative to a component list $X_J$\;
Set $S\gets S\cup L$\;
\KwRet{The strong set $S$ and component lists $X_J$ with $B_J=B(c_J,R_J)$}\;
\end{algorithm}

\begin{figure}[ht]
\centering
\begin{tikzpicture}[scale=1.15,
   pt/.style={circle,fill=black!75,inner sep=1.5pt},
   ctr/.style={circle,draw=blue,thick,inner sep=1.4pt},
   cand/.style={draw=blue!45,dashed,fill=blue!5},
   tight/.style={draw=blue,thick,fill=blue!10},
   faded/.style={draw=black!18,dashed},
   single/.style={draw=red!70,thick},
   drop/.style={draw=red!70,thick,dashed},
   capt/.style={font=\footnotesize,text width=4.6cm,align=center}]

\begin{scope}
  \draw[cand] (1.25,1.7) circle (0.8);
  \draw[cand] (1.25,1.7) circle (0.45);
  \draw[cand] (1.9,0.85) circle (0.8);
  \draw[cand] (1.9,0.85) circle (0.5);
  \draw[cand] (2.95,2.5) circle (0.82);
  \draw[cand] (2.95,2.5) circle (0.5);
  \draw[cand] (0.35,0.7) circle (0.58);
  \draw[cand] (0.35,0.7) circle (0.4);
  \foreach \x/\y in {0.3/2.2, 1.0/2.4, 1.95/2.25, 0.5/1.5, 2.2/1.72, 3.0/1.95, 2.6/2.75,
                     1.35/0.95, 2.95/0.6, 1.6/0.45, 2.6/0.1, 0.9/0.05, 0.55/0.45}{
     \node[pt] at (\x,\y) {};
  }
  \foreach \x/\y in {1.25/1.7, 1.9/0.85, 2.95/2.5, 0.35/0.7}{
     \node[ctr] at (\x,\y) {};
  }
  \node[capt] at (1.6,-0.75) {the candidate balls after sampling};
\end{scope}

\draw[-{Stealth},thick,black!60] (3.5,1.1) -- (4.6,1.1);

\begin{scope}[shift={(5.0,0)}]
  \draw[tight] (1.25,1.7) circle (0.8);
  \draw[tight] (1.9,0.85) circle (0.8);
  \draw[tight] (2.95,2.5) circle (0.82);
  \draw[tight] (0.35,0.7) circle (0.58);
  \draw[faded] (1.25,1.7) circle (0.45);
  \draw[faded] (1.9,0.85) circle (0.5);
  \draw[faded] (2.95,2.5) circle (0.5);
  \draw[faded] (0.35,0.7) circle (0.4);
  \foreach \x/\y in {0.3/2.2, 1.0/2.4, 1.95/2.25, 0.5/1.5, 2.2/1.72, 3.0/1.95, 2.6/2.75,
                     1.35/0.95, 2.95/0.6, 1.6/0.45, 2.6/0.1, 0.9/0.05, 0.55/0.45}{
     \node[pt] at (\x,\y) {};
  }
  \foreach \x/\y in {0.3/2.2, 1.95/2.25, 2.2/1.72, 2.95/0.6, 2.6/0.1, 0.9/0.05}{
     \draw[single] (\x,\y) circle (0.16);
  }
  \foreach \x/\y in {1.25/1.7, 1.9/0.85, 2.95/2.5, 0.35/0.7}{
     \node[ctr] at (\x,\y) {};
  }
  \node[capt] at (1.6,-0.75) {grow until some candidates are tight; uncovered points stay singletons};
\end{scope}

\draw[-{Stealth},thick,black!60] (8.5,1.1) -- (9.6,1.1);

\begin{scope}[shift={(10.0,0)}]
  \draw[drop] (1.25,1.7) circle (0.8);
  \draw[red!70,thick] (1.05,1.5) -- (1.45,1.9);
  \draw[red!70,thick] (1.05,1.9) -- (1.45,1.5);
  \draw[tight] (1.9,0.85) circle (0.8);
  \draw[tight] (2.95,2.5) circle (0.82);
  \draw[tight] (0.35,0.7) circle (0.58);
  \foreach \x/\y in {0.3/2.2, 1.0/2.4, 1.95/2.25, 0.5/1.5, 2.2/1.72, 3.0/1.95, 2.6/2.75,
                     1.35/0.95, 2.95/0.6, 1.6/0.45, 2.6/0.1, 0.9/0.05, 0.55/0.45}{
     \node[pt] at (\x,\y) {};
  }
  \foreach \x/\y in {0.3/2.2, 1.95/2.25, 2.2/1.72, 2.95/0.6, 2.6/0.1, 0.9/0.05}{
     \draw[single] (\x,\y) circle (0.16);
  }
  \foreach \x/\y in {1.9/0.85, 2.95/2.5, 0.35/0.7}{
     \node[ctr] at (\x,\y) {};
  }
  \node[capt] at (1.6,-0.75) {select: keep disjoint tight balls, drop the rest};
\end{scope}

\end{tikzpicture}
\caption{Rounding in three stages: candidates, the tight candidates produced by growth (solid) and the singletons (red), and the pairwise disjoint balls kept by selection.
A centre may have candidates of several radii.}
\label{fig:rounding}
\end{figure}
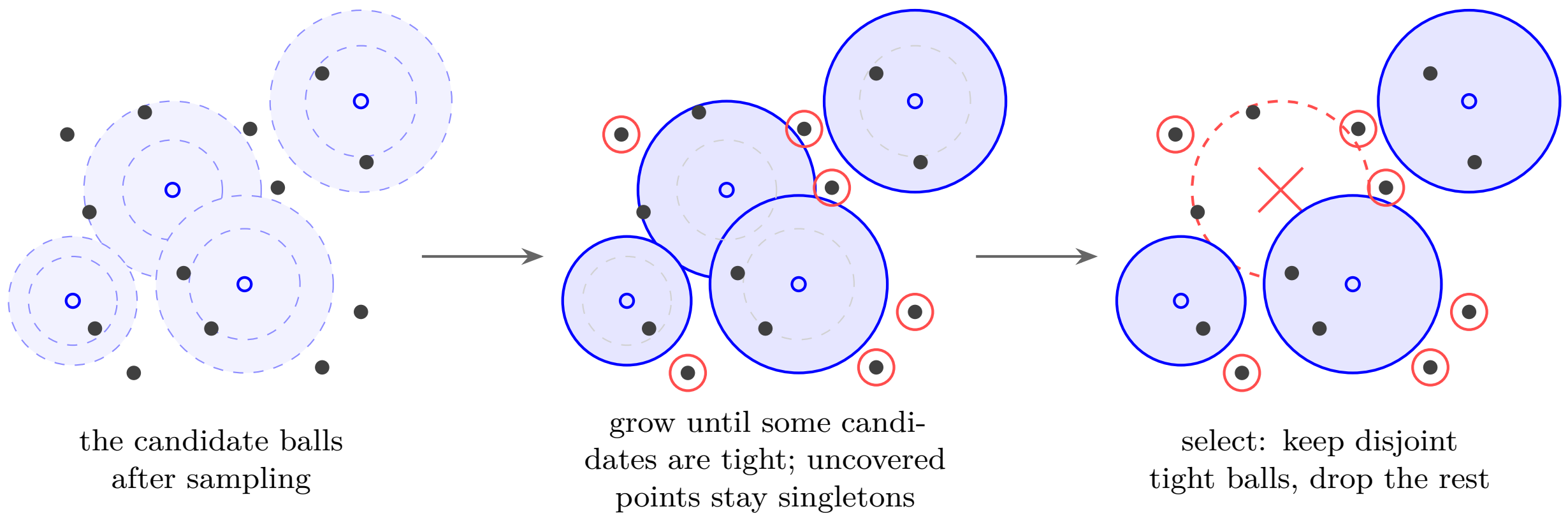
\begin{figure}[ht]
\centering
\begin{tikzpicture}[scale=0.75,
   pt/.style={circle,fill=black!75,inner sep=1.4pt},
   ctr/.style={circle,fill=blue,inner sep=1.5pt},
   root/.style={circle,fill=red,inner sep=2pt},
   core/.style={draw=blue,thick,fill=blue!10},
   infl/.style={draw=blue!55,dashed},
   ballJ/.style={draw=green!55!black,thick,fill=green!55!black,fill opacity=0.05},
   lbl/.style={font=\footnotesize},
   capt/.style={font=\footnotesize,text width=5.2cm,align=center,anchor=north}]

\begin{scope}
  \draw[infl] (1.9,0.85) circle (1.0);
  \draw[infl] (2.7,2.35) circle (1.3);
  \draw[infl] (-0.9,0.7) circle (0.5);
  \draw[core] (1.9,0.85) circle (0.62);
  \draw[core] (2.7,2.35) circle (0.8);
  \draw[core] (-0.9,0.7) circle (0.32);
  \foreach \x/\y in {1.6/1.3, 2.2/0.4, 1.45/0.5, 2.5/1.1, 2.35/2.9, 3.2/2.0,
                     2.2/2.0, 3.3/2.75, -0.75/0.9, 2.9/1.55}{
     \node[pt] at (\x,\y) {};
  }
  \foreach \x/\y in {1.9/0.85, 2.7/2.35, -0.9/0.7}{
     \node[ctr] at (\x,\y) {};
  }
  \node[capt] at (1.3,-0.6) {the selected balls inflate, and some of them overlap};
\end{scope}

\draw[-{Stealth},thick,black!60] (4.4,1.5) -- (5.6,1.5);

\begin{scope}[shift={(7.7,0)}]
  \draw[ballJ] (2.7,2.35) circle (2.7);
  \draw[infl] (1.9,0.85) circle (1.0);
  \draw[infl] (2.7,2.35) circle (1.3);
  \draw[infl] (-0.9,0.7) circle (0.5);
  \draw[red!80,thick] (1.9,0.85) -- (2.7,2.35);
  \foreach \x/\y in {1.6/1.3, 2.2/0.4, 1.45/0.5, 2.5/1.1, 2.35/2.9, 3.2/2.0,
                     2.2/2.0, 3.3/2.75, -0.75/0.9, 2.9/1.55}{
     \node[pt] at (\x,\y) {};
  }
  \node[ctr] at (1.9,0.85) {};
  \node[ctr] at (-0.9,0.7) {};
  \node[root] at (2.7,2.35) {};
  \node[lbl,text=blue] at (1.72,0.6) {$s_i$};
  \node[lbl,text=red] at (3.15,2.5) {$c_J$};
  \node[lbl,text=green!45!black] at (3.875,4.385) {$B_J$};
  \node[lbl,text=black!60,anchor=west] at (4.85,0.3) {large};
  \node[lbl,text=black!60] at (-0.9,1.45) {small};
  \node[capt] at (1.9,-0.6) {overlapping balls merge into one component with a root; a lone ball with few points is a small component};
\end{scope}

\end{tikzpicture}
\caption{Merging the selected balls into components.
Selected balls are joined using the centre-distance threshold $7(R_i+R_j)$; the dashed inflated balls may overlap. Each graph component has a root $c_J$, and its ball $B_J$ (green) contains its inflated balls (\Cref{lem:components}).}
\label{fig:merging}
\end{figure}
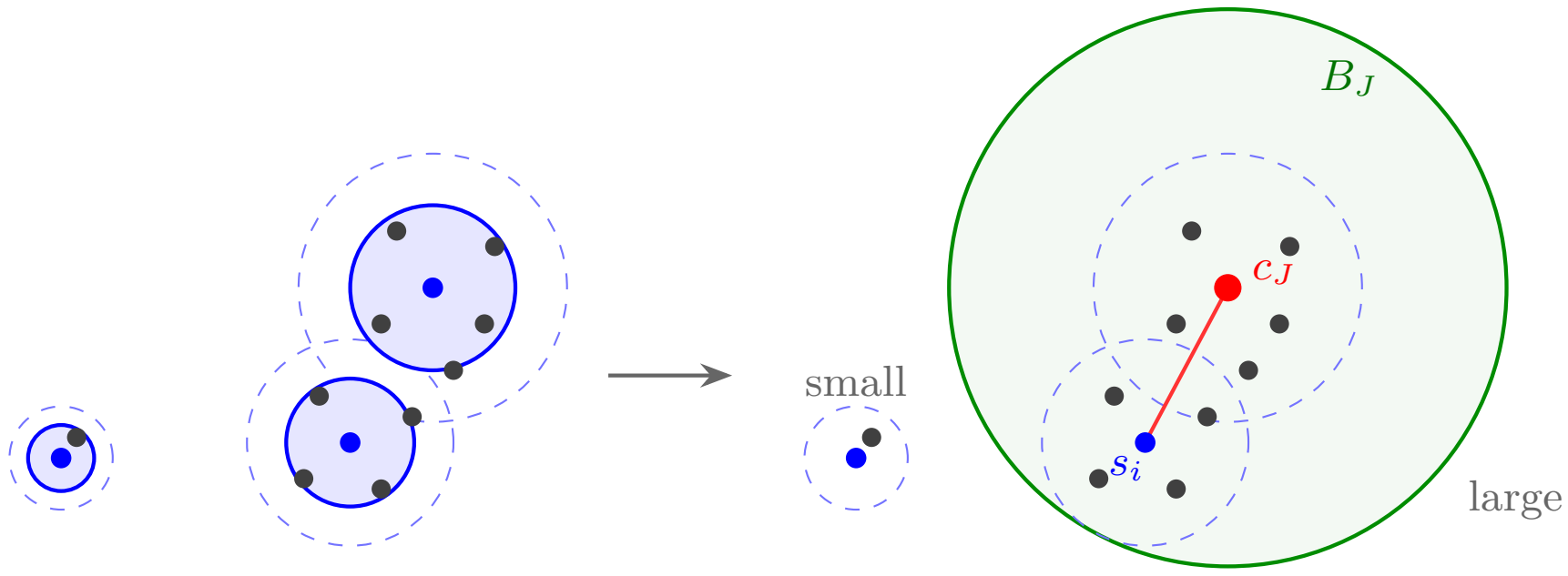

\subsubsection{Bounds from the packing LP}

\begin{lemma}\label{lem:feasible}
The growth phase ends with a feasible solution of \eqref{eq:packing}. In particular,
\begin{equation}\label{eq:yvalue}
 y(X\setminus S_{\mathrm{in}})\le\OPTLP,
\end{equation}
where $S_{\mathrm{in}}$ is the strong set at the start of rounding and, recall, $\OPTLP$ is the optimum LP solution value for the final set of candidates $\mathcal C$. 

\end{lemma}
\begin{proof}
Every variable increases continuously until it freezes. A candidate constraint cannot be violated: when it first becomes tight, all variables in that candidate are frozen. The same is true for each cap constraint. Each event freezes a previously unfrozen variable, so there are at most $|X\setminus S_{\mathrm{in}}|$ effective events. The resulting vector is feasible and therefore has objective at most $\OPTLP$.
\end{proof}

\begin{lemma}\label{lem:singletons}
On the event of \Cref{lem:sampling}, if $\lambda\ge\OPT/(2k)$, then $|L|<51kh$. Thus, for $c\ge51$, rounding does not abort at its first check, and querying $L$ uses $O(k\log n)$ strong queries.
\end{lemma}
\begin{proof}
Every point of $L$ has weight exactly $\lambda/h$, so \Cref{lem:feasible,lem:sampling} yield
\[
 |L|\frac\lambda h=y(L)\le\OPTLP
 <17(\OPT+k\lambda)\le51k\lambda
\]
on the event of \Cref{lem:sampling}. Thus $|L|<51kh$, below $ckh$ for $c\ge51$.
\end{proof}

\begin{lemma}\label{lem:accepted}
On the events of \Cref{lem:fewcorrupt,lem:sampling}, if $\lambda\ge\OPT/(2k)$, then the selected family $Z'$ satisfies $m<102k$ and $\sum_{i=1}^mR_i<51k\lambda$. Thus, for $c\ge102$, rounding does not abort at its second check.
\end{lemma}
\begin{proof}
Selected balls $B(s_i,R_i)$ are pairwise disjoint. By \Cref{lem:proxy}, deleting $E_{s_i}$ from $A(s_i,R_i)$ leaves a subset of $B(s_i,R_i)$; the sets $A(s_i,R_i)\setminus E_{s_i}$ are therefore pairwise disjoint. Moreover, each selected candidate is tight, so \eqref{eq:badweight} implies
\[
 y(A(s_i,R_i)\setminus E_{s_i})
 \ge R_i+\lambda-y(E_{s_i})>R_i+\frac\lambda2.
\]
Summing over the disjoint sets and using \eqref{eq:yvalue} gives
\[
 \sum_{i=1}^m\left(R_i+\frac\lambda2\right)<\OPTLP
 <17(\OPT+k\lambda)\le51k\lambda.
\]
Thus $\sum_iR_i<51k\lambda$ and $m<102k$. In particular, rounding does not abort at its second check when $c\ge102$.
\end{proof}

\subsubsection{From candidates to true balls}

Although membership in $A(s,R)$ is determined by estimates, membership in the actual ball $B(s,R)$ is not known for unqueried points. The sets $E_s$ of points corrupt for $W_s$ bridge this gap.

\begin{figure}[ht]
\centering
\begin{tikzpicture}[scale=0.68,
   pt/.style={circle,fill=black!75,inner sep=1.8pt},
   ctr/.style={circle,fill=blue,inner sep=2.1pt},
   lbl/.style={font=\footnotesize}]
  \draw[fill=blue!5,draw=blue!45,dashed] (0,0) circle (3.6);
  \draw[fill=black!7,draw=black!45] (2.0,0) circle (1.2);
  \draw[fill=blue!14,fill opacity=0.7,draw=blue!70] (0,0) circle (1.2);
  \node[ctr] (si) at (0,0) {};
  \node[lbl,text=blue] at (-0.3,-0.33) {$s_i$};
  \node[ctr] (s) at (2.0,0) {};
  \node[lbl,text=blue] at (2.02,-0.36) {$s$};
  \node[pt] (x) at (3.2,0) {};
  \node[lbl] at (3.1,-0.42) {$x$};
  \draw[blue,-{Latex[length=1.5mm]}] (0,0) -- (100:1.2);
  \node[lbl,text=blue] at (0.3,0.6) {$R_i$};
  \draw[blue!60,-{Latex[length=1.5mm]}] (0,0) -- (152:3.6);
  \node[lbl,text=blue] at (-2.1,1.55) {$3R_i$};
  \draw[black!65,{Latex[length=1.3mm]}-{Latex[length=1.3mm]}] (0,-1.75) -- (2.0,-1.75);
  \node[lbl] at (1.0,-2.25) {$\le R+R_i$};
  \draw[black!65,{Latex[length=1.3mm]}-{Latex[length=1.3mm]}] (2.0,0.42) -- (3.2,0.42);
  \node[lbl] at (2.6,0.9) {$\le R$};
\end{tikzpicture}
\caption{
The candidate $A(s,R)$ has radius $R\le R_i$ and satisfies $d(s,s_i)\le R+R_i$.
It is assigned to the selected candidate $A(s_i,R_i)$.}
\label{fig:inflate}
\end{figure}
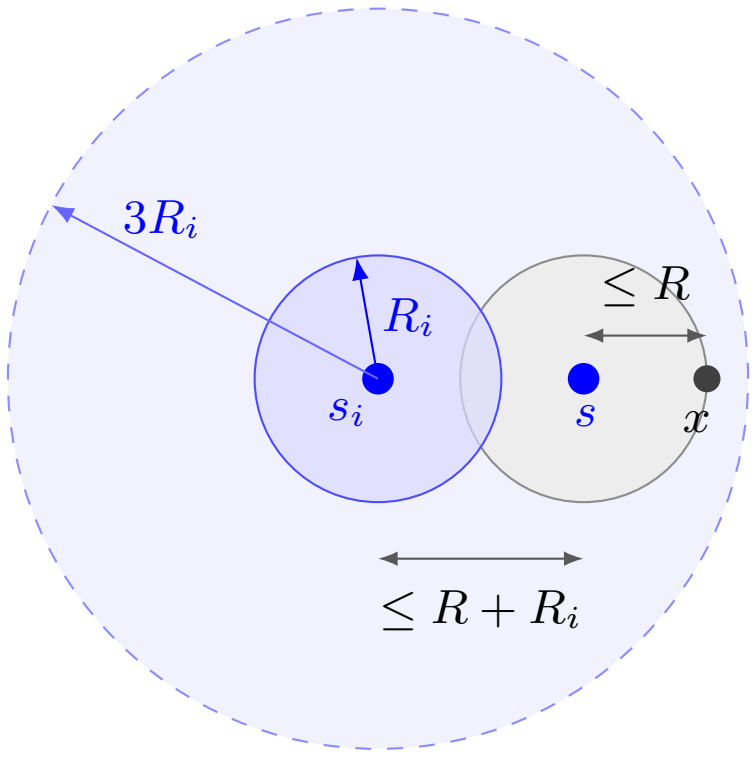

\begin{lemma}\label{lem:inflate}
Whenever rounding does not abort, every point $x\notin E$ that is not strongly queried lies in $B(s_i,3R_i)$ for the selected candidate $i$ through which it is assigned. On the event of \Cref{lem:globalcorrupt}, we also have $|E|\le2ckh=O(kh)$.
\end{lemma}
\begin{proof}
\Cref{lem:globalcorrupt} gives
\[
 |E|\le2ckh=O(k\log n).
\]

Let $(A(s,R),s,R)$ be the pruned candidate through which $x$ is assigned. It contains the candidate that originally froze $x$, so $x\in A(s,R)$. Since $x\notin E_s$, \Cref{lem:proxy} gives $d(s,x)\le R$.

If this candidate is selected, the claim follows with $s_i=s$ and $R_i=R$. Otherwise it was discarded in favour of a previously selected candidate $i$ with $R\le R_i$ and $d(s,s_i)\le R+R_i$ (see \Cref{fig:inflate}). Thus
\[
 d(s_i,x)\le d(s_i,s)+d(s,x)
 \le R_i+2R\le3R_i.
\]
\end{proof}

For a graph component $J$, define the (analytical, not necessarily computable) union
\begin{equation}\label{eq:union}
 B'_J:=\bigcup_{i\in J}B(s_i,5R_i).
\end{equation}
The inflation factor $5$ leaves enough slack for subsequent estimated-distance tests, which may overestimate true distances by $2R_i$. The joining threshold $7(R_i+R_j)$ similarly absorbs the estimation slack between components.

\begin{figure}[ht]
\centering
\begin{tikzpicture}[scale=0.62,
   pt/.style={circle,fill=black!75,inner sep=1.7pt},
   ctr/.style={circle,fill=blue,inner sep=1.9pt},
   root/.style={circle,fill=red,inner sep=2.1pt},
   lbl/.style={font=\footnotesize}]
  \draw[fill=blue!8,draw=blue!50] (0,0) circle (1.3);
  \draw[fill=blue!8,draw=blue!50] (2.5,0.35) circle (1.45);
  \draw[fill=blue!8,draw=blue!50] (4.9,-0.2) circle (1.2);
  \node[root] at (0,0) {};
  \node[lbl,text=red] at (-0.1,-0.45) {$c_J$};
  \node[ctr] at (2.5,0.35) {};
  \node[lbl,text=blue] at (2.52,-0.1) {$s_2$};
  \node[ctr] at (4.9,-0.2) {};
  \node[lbl,text=blue] at (4.92,-0.65) {$s_3$};
  \node[pt] at (6.1,-0.2) {};
  \node[lbl] at (6.5,-0.2) {$y$};
  \draw[black!75,dashed] (4.9,-0.2) -- (6.1,-0.2);
  \node[lbl] at (5.5,0.3) {$5R_3$};
  \draw[black!70,{Latex[length=1.4mm]}-{Latex[length=1.4mm]}] (0,-2.1) -- (2.5,-2.1);
  \node[lbl] at (1.25,-2.65) {$\le7(R_1+R_2)$};
  \draw[black!70,{Latex[length=1.4mm]}-{Latex[length=1.4mm]}] (2.5,-3.15) -- (4.9,-3.15);
  \node[lbl] at (3.7,-3.7) {$\le7(R_2+R_3)$};
  \draw[green!55!black,thick] (-32:7.9) arc(-32:32:7.9);
  \draw[green!55!black,-{Latex[length=1.6mm]}] (0,0) -- (20:7.9);
  \node[lbl,text=green!45!black] at (3.1,2.55) {$R_J=14\sum_{i\in J}R_i$};
  \node[lbl,text=green!45!black] at (8.3,-1.2) {$B_J$};
\end{tikzpicture}
\caption{A component and the walk across it.
A walk from the root to $s_3$ pays $7(R_i+R_j)$ at every step, and the last step to $y$ pays $5R_3$.
This puts $y$ within $R_J$ of the root.}
\label{fig:component}
\end{figure}
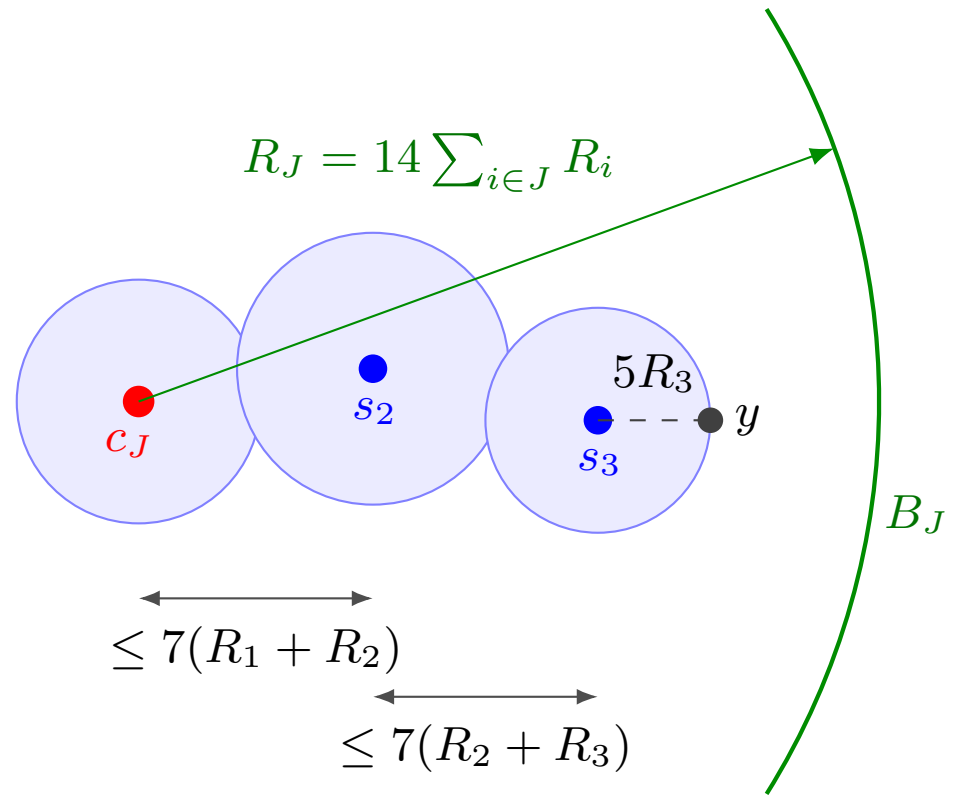

\begin{lemma}\label{lem:components}
There are at most $m$ graph components, and for each component $J$, $B'_J\subseteq B_J$. Moreover,
\[
 \sum_JR_J=14\sum_{i=1}^mR_i.
\]
\end{lemma}
\begin{proof}
The bound on the number of components is immediate, and the sum follows from their partition of $\{1,\ldots,m\}$. Let $x\in B(s_i,5R_i)$ for $i\in J$. Choose a simple path from the root $c_J=s_{i_0}$ to $s_i=s_{i_t}$ in the component graph (see \Cref{fig:component}). By construction,
\[
 d(c_J,s_i)\le7\sum_{\ell=0}^{t-1}(R_{i_\ell}+R_{i_{\ell+1}})
 \le14\sum_{j\in J}R_j-7R_i.
\]
The last bound is also valid when $t=0$. Therefore
\[
 d(c_J,x)\le d(c_J,s_i)+5R_i\le14\sum_{j\in J}R_j=R_J,
\]
which proves $B'_J\subseteq B_J$.
\end{proof}

\Cref{lem:list} compares the lists with these unions.

\begin{lemma}\label{lem:list}
Whenever rounding does not abort, for each component $J$, after the singleton queries have been added to $S$,
\begin{equation}\label{eq:list}
 X_J\setminus E\subseteq B'_J.
\end{equation}
\end{lemma}
\begin{proof}
If $x\in X_J\setminus E$, its selected representative $i$ belongs to $J$, and \Cref{lem:inflate} gives $x\in B(s_i,3R_i)\subseteq B'_J$.
\end{proof}

\begin{proof}[Proof of \Cref{lem:rounding}]
On the events of \Cref{lem:fewcorrupt,lem:globalcorrupt,lem:sampling}, \Cref{lem:singletons,lem:accepted} prevent both aborts and establish the query and cardinality bounds. Each point left unqueried was frozen through a candidate, so it belongs to exactly one component list. By \Cref{lem:components,lem:accepted},
\[
 \sum_JR_J=14\sum_iR_i=O(k\lambda).
\]
Finally, \Cref{lem:list,lem:components} give $X_J\setminus E\subseteq B'_J\subseteq B_J$, and \Cref{lem:inflate} bounds $|E|$. Every root $c_J$ is the centre of a sampled candidate and is therefore strongly queried.
\end{proof}

\subsection{Repair}\label{sec:repair}

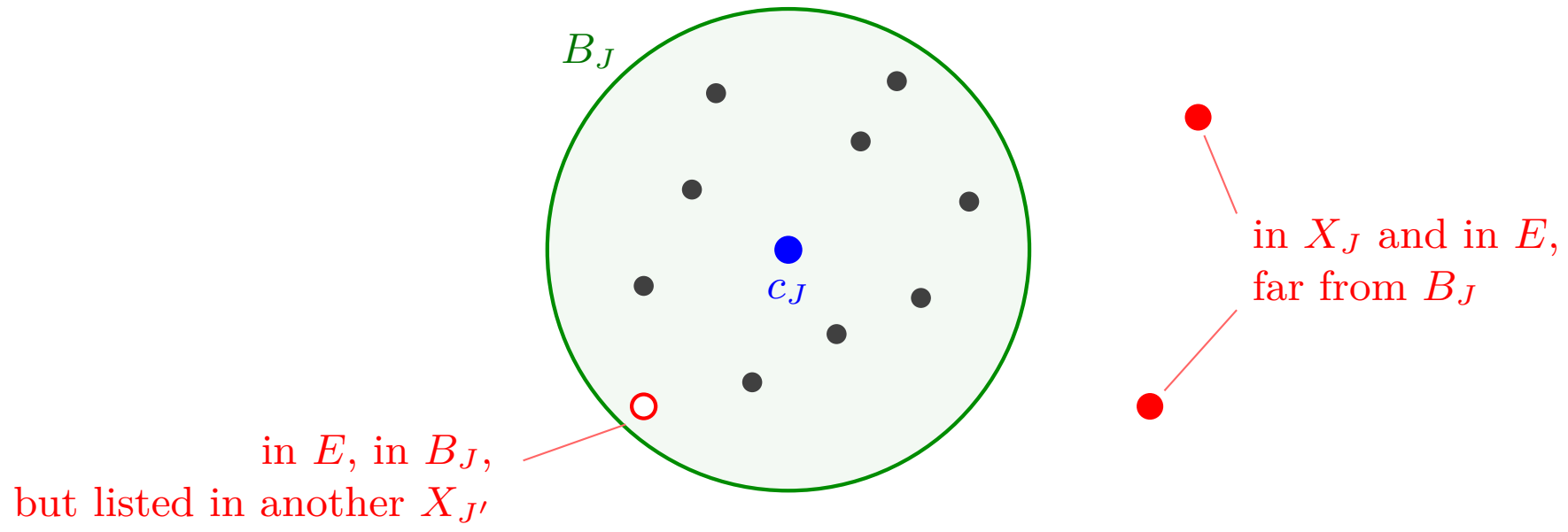
\begin{figure}[ht]
\centering
\begin{tikzpicture}[scale=0.9,
   pt/.style={circle,fill=black!75,inner sep=1.5pt},
   bad/.style={circle,fill=red,inner sep=2pt},
   away/.style={circle,draw=red,thick,fill=white,inner sep=1.8pt},
   root/.style={circle,fill=blue,inner sep=2.1pt},
   lbl/.style={font=\footnotesize}]
  \draw[draw=green!55!black,thick,fill=green!55!black,fill opacity=0.05] (0,0) circle (2.0);
  \node[root] at (0,0) {};
  \node[lbl,text=blue] at (0,-0.35) {$c_J$};
  \node[lbl,text=green!45!black] at (-1.65,1.65) {$B_J$};
  \foreach \x/\y in {0.6/0.9, -0.8/0.5, 1.1/-0.4, -0.3/-1.1, 0.4/-0.7,
                     -1.2/-0.3, 0.9/1.4, -0.6/1.3, 1.5/0.4}{
     \node[pt] at (\x,\y) {};
  }
  \node[bad] (b1) at (3.4,1.1) {};
  \node[bad] (b2) at (3.0,-1.3) {};
  \node[lbl,text=red,align=left,anchor=west] at (3.7,-0.1) {in $X_J$ and in $E$,\\ far from $B_J$};
  \node[away] (a1) at (-1.2,-1.3) {};
  \node[lbl,text=red,align=right,anchor=east] at (-2.3,-1.9) {in $E$, in $B_J$,\\ but listed in another $X_{J'}$};
  \draw[red!60,thin] (-1.35,-1.45) -- (-2.2,-1.75);
  \draw[red!60,thin] (3.45,0.95) -- (3.72,0.3);
  \draw[red!60,thin] (3.12,-1.17) -- (3.72,-0.5);
\end{tikzpicture}
\caption{
On the event of \Cref{lem:globalcorrupt}, the black points in $X_J\setminus E$ lie in the ball $B_J$ (\Cref{lem:inflate,lem:components}).
The red points lie in $E$: they may sit in $X_J$ far from $B_J$, or lie in $B_J$ and be listed in another component.
Solid red points are in $X_J$; the hollow one is not.
The algorithm does not know which points are in $E$.}
\label{fig:wronglist}
\end{figure}

Rounding produces a small family of balls, but the lists $X_J$ were assembled using estimated distances. On the event of \Cref{lem:globalcorrupt}, if rounding does not abort, every unqueried point outside $E$ is assigned to a ball that contains it by \Cref{lem:inflate,lem:components}; a point in $E$ may be assigned incorrectly or omitted from a component to which it genuinely belongs (see \Cref{fig:wronglist}). The repair step makes these assignments reliable. It first queries points with demonstrably inconsistent estimates, then removes small components by querying all their points, and finally tests the remaining points using new, uniformly sampled witnesses from large components.

A summary of what we prove in this section is captured by the following lemma.
\begin{lemma}\label{lem:repair}
With high probability, whenever rounding does not abort, repair does not abort either and returns an updated strong set $S$ and component lists $X_J$ satisfying
\begin{enumerate}
\item $X_J\subseteq B(c_J,3R_J)$ for every component $J$;
\item $S\cup\bigcup_JX_J=X$; and
\item repair makes $O(k\log n)$ additional strong queries.
\end{enumerate}
The final lists need not be disjoint.
\end{lemma}

\subsubsection{Corrupt points and estimation slack}

Unlike $E$, the set $E_J$ involves only selected candidates within $J$. On the event of \Cref{lem:fewcorrupt}, $|E_{s_i}|<h/2$.
On the event of \Cref{lem:globalcorrupt},
\begin{equation}\label{eq:Esize}
 |E|\le2ckh.
\end{equation}

Every selected candidate $(A(s_i,R_i),s_i,R_i)$ satisfies $r_{s_i}\le R_i$: the candidate was tight and therefore contained an unqueried point of positive weight, whose estimate, by \eqref{eq:score}, was at least $r_{s_i}$. Thus \Cref{lem:proxy} gives the following useful estimate for every $x\notin E_{s_i}$:
\begin{equation}\label{eq:sandwich}
 d(s_i,x)\le\md(x,s_i)\le d(s_i,x)+2R_i.
\end{equation}
In particular, a point $x\notin E_{s_i}$ within $3R_i$ of $s_i$ has estimate at most $5R_i$, and if it is within $5R_i$, its estimate is at most $7R_i$. These are the two thresholds used in repair.

\subsubsection{The repair procedure}

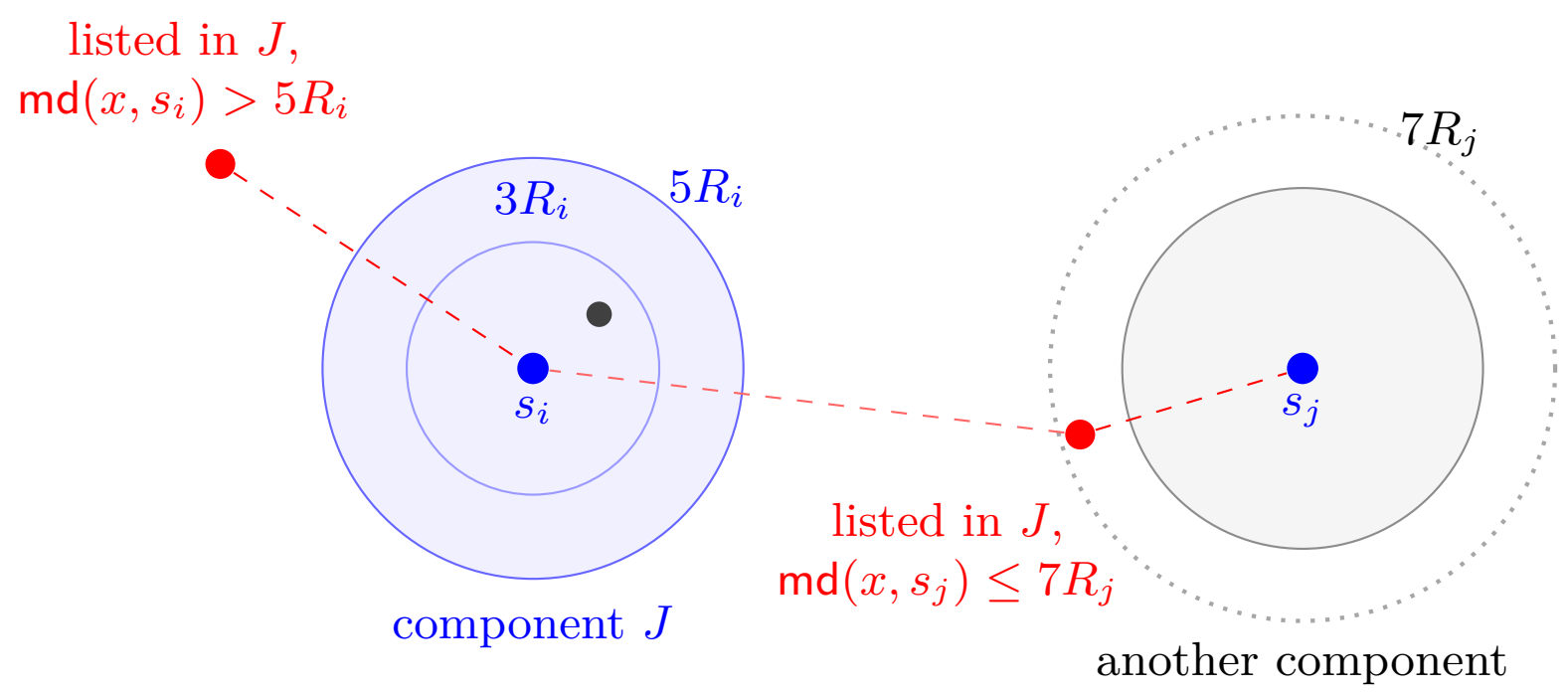
\begin{figure}[ht]
\centering
\begin{tikzpicture}[scale=0.8,
   pt/.style={circle,fill=black!75,inner sep=1.7pt},
   bad/.style={circle,fill=red,inner sep=2pt},
   ctr/.style={circle,fill=blue,inner sep=2.1pt},
   lbl/.style={font=\footnotesize}]
  \draw[fill=blue!6,draw=blue!60] (0,0) circle (1.75);
  \draw[draw=blue!40] (0,0) circle (1.05);
  \node[ctr] (si) at (0,0) {};
  \node[lbl,text=blue] at (0,-0.35) {$s_i$};
  \node[lbl,text=blue] at (0,1.4) {$3R_i$};
  \node[lbl,text=blue] at (1.45,1.5) {$5R_i$};
  \node[lbl,text=blue] at (0,-2.15) {component $J$};
  \draw[fill=black!4,draw=black!45] (6.4,0) circle (1.5);
  \draw[draw=black!35,dotted,thick] (6.4,0) circle (2.1);
  \node[ctr] (sj) at (6.4,0) {};
  \node[lbl,text=blue] at (6.4,-0.35) {$s_j$};
  \node[lbl] at (7.55,1.95) {$7R_j$};
  \node[lbl] at (6.4,-2.45) {another component};
  \node[pt] at (0.55,0.45) {};
  \node[bad] (x2) at (-2.6,1.7) {};
  \node[lbl,text=red,align=center] at (-2.9,2.45) {listed in $J$,\\ $\md(x,s_i)>5R_i$};
  \draw[red,dashed] (x2) -- (si);
  \node[bad] (x3) at (4.55,-0.55) {};
  \node[lbl,text=red,align=center] at (3.45,-1.55) {listed in $J$,\\ $\md(x,s_j)\le7R_j$};
  \draw[red,dashed] (x3) -- (sj);
  \draw[red!60,dashed] (x3) -- (si);
\end{tikzpicture}
\caption{The two conflict checks.
Both red points are listed in $J$, and both enter the set $D$.
The black point does not.}
\label{fig:conflict}
\end{figure}

First, let $D$ contain every currently unqueried point $x$ whose assigned selected representative $i$ violates $\md(x,s_i)\le5R_i$, or for which $\md(x,s_j)\le7R_j$ for some selected candidate $j$ in a \emph{different} component. Query all of $D$ and denote the resulting strong set by $S_D$. \Cref{fig:conflict} shows the two ways a point enters $D$. On the event of \Cref{lem:globalcorrupt}, \Cref{lem:conflict} shows that $D\subseteq E$.

For each component $J$, remove strongly queried points from its old list and classify it as \emph{small} if
\[
 |X_J|<2c\bigl(|S_D\cap B'_J|+|J|h\bigr),
\]
and \emph{large} otherwise. The expression $|S_D\cap B'_J|$ is computable, since all selected centres and all points of $S_D$ are strongly queried. Classify all components against the \emph{same} strong set $S_D$ before making further queries. Strongly query and empty the lists of small components.

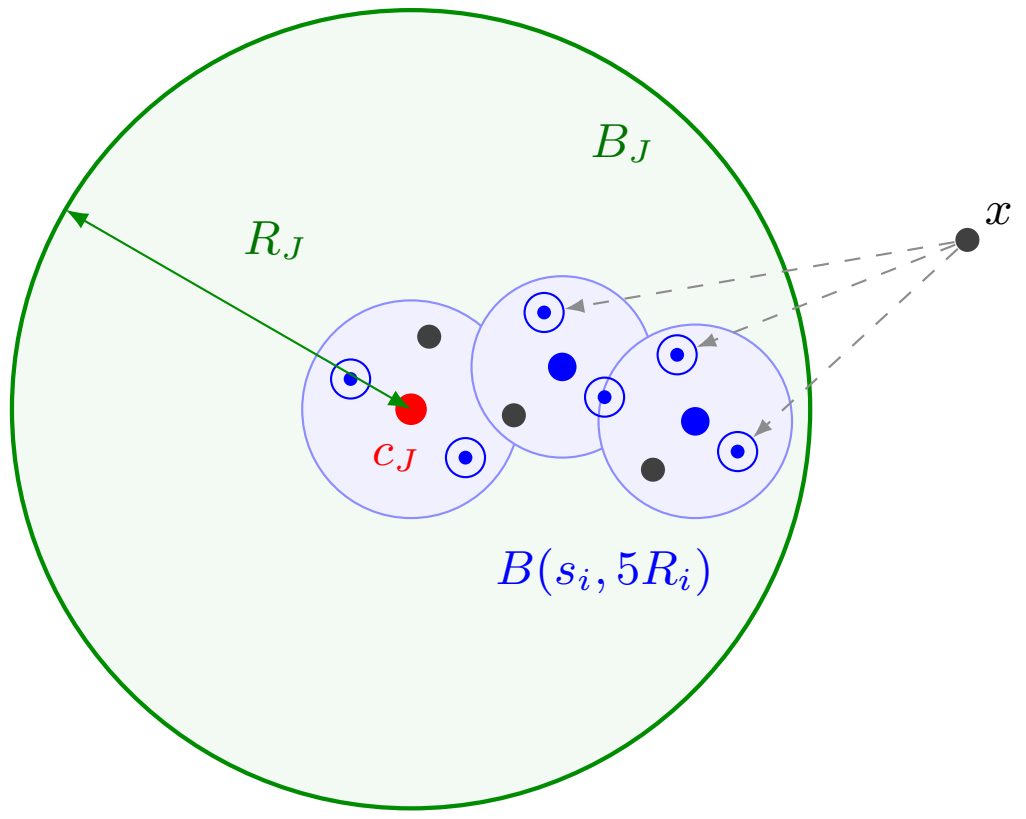
\begin{figure}[ht]
\centering
\begin{tikzpicture}[scale=0.8,
   pt/.style={circle,fill=black!75,inner sep=1.6pt},
   ctr/.style={circle,fill=blue,inner sep=1.9pt},
   root/.style={circle,fill=red,inner sep=2.1pt},
   wit/.style={circle,draw=blue,inner sep=2.6pt,path picture={\fill[blue] (path picture bounding box.center) circle (1.3pt);}},
   lbl/.style={font=\footnotesize}]
  \draw[draw=green!55!black,thick,fill=green!55!black,fill opacity=0.05] (0,0) circle (3.3);
  \node[lbl,text=green!45!black] at (1.75,2.2) {$B_J$};
  \draw[fill=blue!6,draw=blue!45] (0,0) circle (0.9);
  \draw[fill=blue!6,draw=blue!45] (1.25,0.35) circle (0.75);
  \draw[fill=blue!6,draw=blue!45] (2.35,-0.1) circle (0.8);
  \node[root] (c) at (0,0) {};
  \node[lbl,text=red] at (-0.12,-0.4) {$c_J$};
  \node[ctr] at (1.25,0.35) {};
  \node[ctr] at (2.35,-0.1) {};
  \node[lbl,text=blue] at (1.6,-1.35) {$B(s_i,5R_i)$};
  \node[wit] (wa) at (0.45,-0.4) {};
  \node[wit] (wb) at (-0.5,0.25) {};
  \node[wit] (wc) at (1.1,0.8) {};
  \node[wit] (wd) at (1.6,0.1) {};
  \node[wit] (we) at (2.2,0.45) {};
  \node[wit] (wf) at (2.7,-0.35) {};
  \node[pt] at (0.15,0.6) {};
  \node[pt] at (2.0,-0.5) {};
  \node[pt] at (0.85,-0.05) {};
  \draw[green!55!black,{Latex[length=1.5mm]}-{Latex[length=1.5mm]}] (0,0) -- (150:3.3);
  \node[lbl,text=green!45!black] at (-1.12,1.40) {$R_J$};
  \node[pt] (x) at (4.6,1.4) {};
  \node[lbl] at (4.86,1.62) {$x$};
  \foreach \w in {wc,we,wf}{
     \draw[black!45,thin,dashed,-{Latex[length=1.3mm]}] (x) -- (\w);
  }
\end{tikzpicture}
\caption{A large component with root $c_J$, where every point of $B'_J$ is within $R_J$ of the root (\Cref{lem:components}).
The new witnesses $W_J$ (ringed blue) are drawn from the list $X_J$.
A point $x$ joins $X_J$ when the median of $\Tilde d(x,w)+d(c_J,w)$ over $w\in W_J$ is at most $3R_J$.}
\label{fig:largecomponent}
\end{figure}

For every large component, select a fresh set $W_J$ of $h$ distinct points uniformly from its current list, and strongly query them. The list-size threshold ensures that such a sample exists. For a point $x$ not yet strongly queried, define its \emph{component estimate}
\begin{equation}\label{eq:mdJ}
 \md(x,J):=\operatorname{median}_{w\in W_J}
 \bigl(\Tilde d(x,w)+d(c_J,w)\bigr).
\end{equation}
Recompute the list of each large component as $\{x\in X\setminus S:\md(x,J)\le3R_J\}$. \Cref{fig:largecomponent} illustrates this test. After these lists are computed, strongly query every point belonging to none of them. Unlike the initial partition, the recomputed lists may overlap; this is harmless because a final assignment may select any one qualifying component.

\begin{algorithm}[H]
\caption{\textup{\textsc{Repair}}: repairing component lists.}\label{alg:repair}
\LinesNumbered
\KwIn{The strong set $S$, the component lists $X_J$, their centres $c_J$ and radii $R_J$, and each point's selected representative from rounding}
$D\gets\{x\in X\setminus S:\text{$x$ is assigned through $i$ and $\md(x,s_i)>5R_i$}\}$\;
Add to $D$ any $x\in X\setminus S$ assigned to component $J$ for which $\md(x,s_j)\le7R_j$ for some $j\notin J$\;
\lIf{$|D|>2ckh$}{\textbf{abort}}
Strongly query $D$; set $S\gets S\cup D$ and $S_D\gets S$\;
\For{each component $J$}{
 Set $X_J\gets X_J\setminus S_D$; classify $J$ as small if $|X_J|<2c(|S_D\cap B'_J|+|J|h)$, and large otherwise\;
}
\For{each small component $J$}{
 Strongly query $X_J$; add its points to $S$ and set $X_J\gets\varnothing$\;
}
\For{each large component $J$}{
 Draw $h$ distinct points $W_J$ uniformly from $X_J$; strongly query them and add them to $S$\;
}
\For{each large component $J$}{
 $X_J\gets\{x\in X\setminus S:\md(x,J)\le3R_J\}$\;
}
$U\gets (X\setminus S)\setminus\bigcup_J X_J$\;
\lIf{$|U|>2ckh$}{\textbf{abort}}
Strongly query $U$ and set $S\gets S\cup U$\;
\KwRet{The updated $S$ and lists $X_J$}\;
\end{algorithm}

\subsubsection{Detecting conflicts and controlling list errors}

\begin{lemma}\label{lem:conflict}
On the event of \Cref{lem:globalcorrupt}, whenever rounding does not abort, every point in $D$ belongs to $E$. In particular, $|D|\le2ckh$, so repair does not abort at its first check.
\end{lemma}
\begin{proof}
Take $x\notin E$ and let $i$ be its selected representative in component $J$. By \Cref{lem:inflate}, $d(s_i,x)\le3R_i$, and by \eqref{eq:sandwich}, $\md(x,s_i)\le5R_i$. Thus $x$ fails the first conflict condition.

Suppose it satisfies the second condition for some $j\notin J$. Since $x\notin E_{s_j}$, \eqref{eq:sandwich} implies $d(s_j,x)\le7R_j$. Therefore
\[
 d(s_i,s_j)\le d(s_i,x)+d(x,s_j)
 \le3R_i+7R_j\le7(R_i+R_j).
\]
The rounding graph would connect $i$ and $j$, contradicting $j\notin J$. Thus $x\notin D$, and $D\subseteq E$. The cardinality claim follows from \eqref{eq:Esize}.
\end{proof}

After the conflict queries, the difference between a component list and its true inflated union is controlled by the \emph{local} set $E_J$, whose bound scales with the number of selected centres in that component.

\begin{lemma}\label{lem:listfast}
On every run in which rounding and the first conflict check do not abort, immediately after removing $S_D$ from the old lists, each component $J$ satisfies
\begin{equation}\label{eq:listfast}
 X_J\setminus E_J\subseteq B'_J,
 \qquad B'_J\setminus(S_D\cup E_J)\subseteq X_J.
\end{equation}
\end{lemma}
\begin{proof}
Take $x\in X_J\setminus E_J$ and let $i\in J$ be its selected representative. Since $x\notin S_D$, it was not in $D$, and hence $\md(x,s_i)\le5R_i$. As $x\notin E_{s_i}$, \eqref{eq:sandwich} implies $d(s_i,x)\le5R_i$, so $x\in B'_J$.

Conversely, suppose $x\in B'_J\setminus(S_D\cup E_J)$. Some $i\in J$ satisfies $d(s_i,x)\le5R_i$. By \eqref{eq:sandwich}, $\md(x,s_i)\le7R_i$. Since $x\notin S_D$, it is neither a queried singleton nor a queried conflict, and rounding originally assigned it to some component $J'$. If $J'\ne J$, the second conflict test would have placed $x$ in $D$. Thus $J'=J$, and the point is still in $X_J$.
\end{proof}

\subsubsection{Fresh witnesses in large components}

The witnesses of a large component should lie predominantly inside $B'_J$, and their weak edges should be predominantly honest. The first fact follows from the bound on $|E_J|$; the second requires a uniform concentration bound because the components themselves depend on weak distances.

\begin{lemma}\label{lem:inside}
Conditioned on the event of \Cref{lem:fewcorrupt}, whenever repair reaches the witness-sampling stage, with high probability over the fresh samples, simultaneously for all large components $J$, fewer than $h/16$ witnesses in $W_J$ lie outside $B'_J$.
\end{lemma}
\begin{proof}
By \Cref{lem:listfast}, at most $|E_J|<|J|h/2$ points of the old list $X_J$ lie outside $B'_J$. Since $J$ is large, $|X_J|\ge2c|J|h$. Thus the fraction outside is less than $1/(4c)$. Conditional on the list, $W_J$ is uniform without replacement. By \Cref{lem:hypergeom},
\[
 \Pr[|W_J\setminus B'_J|\ge h/16]
 \le\left(\frac{4e}{c}\right)^{h/16}
 \le2^{-h/16}
\]
for sufficiently large $c$. A union bound over at most $ck$ components gives the claim.

\end{proof}

\begin{lemma}\label{lem:uniform}
With high probability, simultaneously for every $1\le j\le ck$, point $x\in X$, and set $A\subseteq X$ expressible as a union of at most $j$ true metric balls with $|A|\ge cjh$, at most $3|A|/8$ edges from $x$ to $A\setminus\{x\}$ are corrupted.
\end{lemma}
\begin{proof}
Fix $j,x,A$ before drawing the set $\mathsf{Corrupt}$. If $C$ counts the corrupted edges from $x$ to $A\setminus\{x\}$, then $\mathbb E[C]\le|A|/3$. \Cref{lem:hoeffding} (using at most $|A|$ independent edges) yields
\[
 \Pr[C>3|A|/8]
 \le\exp(-|A|/288)
 \le n^{-c^2j/288},
\]
since $|A|\ge cjh\ge c^2j\log n$. For a fixed metric there are at most $n^2$ distinct nonempty balls, as each centre has at most $n$ relevant radii. Thus there are at most $n^{2j}$ unions of at most $j$ balls and $n$ choices of $x$. Taking $c$ large enough that $c^2/288\ge4$, a union bound gives failure probability at most $n^{1-2j}\le n^{-j}$ for fixed $j$. Summing over $1\le j\le ck$ gives at most $2/n$. Crucially, the bound is simultaneous over every such union, so it also applies to balls selected using observed weak distances.
\end{proof}

On the events of \Cref{lem:fewcorrupt,lem:uniform}, apply \Cref{lem:uniform} to $A=B'_J$, a union of $j=|J|\le ck$ balls. By \Cref{lem:listfast} and the definition of a large component,
\begin{equation}\label{eq:bigB}
 |B'_J|\ge|X_J|-|E_J|
 >2c|J|h-\frac{|J|h}{2}
 \ge c|J|h.
\end{equation}
Therefore, on the uniform concentration event, no point has corrupted edges to more than $3|B'_J|/8$ members of $B'_J$.

\begin{lemma}\label{lem:honest}
Conditioned on the events of \Cref{lem:fewcorrupt,lem:uniform}, with high probability over the fresh witness samples, for every large component $J$ and every point $x$ that is unqueried when the component tests are performed, at least $9h/16$ of the edges from $x$ to $W_J$ are honest.
\end{lemma}
\begin{proof}
Fix $J$ and $x$. By \Cref{lem:listfast}, the symmetric difference of the old list $X_J$ and $B'_J$ contains at most
\[
 t_J:=|S_D\cap B'_J|+|E_J|
 <|S_D\cap B'_J|+|J|h
 \le\frac{|X_J|}{2c}
\]
points. At most $3|B'_J|/8$ edges from $x$ to $B'_J$ are corrupted, by \Cref{lem:uniform}. Since $|B'_J|\le|X_J|+t_J$, the number of points in $X_J$ whose edge to $x$ is corrupted is at most
\[
 \frac38(|X_J|+t_J)+t_J
 \le\left(\frac38+\frac1c\right)|X_J|.
\]
The witness set is chosen uniformly \emph{after} the old list is fixed, with fresh randomness. Conditional on the list and the set $\mathsf{Corrupt}$, the expected number of corrupted edges from $x$ to the sampled witnesses is thus at most $(3/8+1/c)h\le13h/32$ for $c\ge32$. The threshold $7h/16$ exceeds this expectation by at least $h/32$. \Cref{lem:hypergeom} implies
\[
 \Pr\left[\text{at least }7h/16\text{ witness edges are corrupted}\right]
 \le\exp(-h/512).
\]
A union bound over at most $ck$ components and $n$ points proves the claim for sufficiently large $c$.
\end{proof}

\begin{definition}\label{def:good}
A witness $w\in W_J$ is \emph{good for $x$} if $w\in B'_J$ and the edge $xw$ is honest.
\end{definition}

\begin{lemma}\label{lem:goodmajority}
With high probability, every large component $J$ and every point $x$ unqueried at the component-testing stage have strictly more than $h/2$ witnesses good for $x$.
\end{lemma}
\begin{proof}
On the events of \Cref{lem:honest,lem:inside}, at least $9h/16$ witnesses have honest edges to $x$, and fewer than $h/16$ are outside $B'_J$. Thus more than $9h/16-h/16=h/2$ witnesses satisfy both conditions.
\end{proof}

\begin{lemma}\label{lem:median}
If more than half the witnesses of a large component $J$ are good for an unqueried point $x$, then
\[
 d(c_J,x)\le\md(x,J)\le d(c_J,x)+2R_J.
\]
Consequently, $x\in B'_J$ implies that the recomputed $X_J$ contains $x$, and membership in that recomputed list implies $x\in B(c_J,3R_J)$.
\end{lemma}
\begin{proof}
For every good witness $w$, \Cref{lem:components} gives $d(c_J,w)\le R_J$, and the triangle inequality gives
\[
 d(c_J,x)\le d(x,w)+d(c_J,w)
 \le d(c_J,x)+2R_J.
\]
Because the edge $xw$ is honest, the middle expression equals $\Tilde d(x,w)+d(c_J,w)$. A strict majority of these expressions lie in the displayed interval, so their median does too. If $x\in B'_J$, then $d(c_J,x)\le R_J$ and $\md(x,J)\le3R_J$. Conversely, if $\md(x,J)\le3R_J$, its lower bound gives $d(c_J,x)\le3R_J$.
\end{proof}

\subsubsection{Completing the repair}

\begin{lemma}\label{lem:residual}
With high probability, every point unqueried after the witnesses have been selected that belongs to no recomputed list $X_J$ lies in $E$. In particular, at most $2ckh$ such points remain, so repair does not abort at its final check.
\end{lemma}
\begin{proof}
Work on the events of \Cref{lem:globalcorrupt,lem:goodmajority}. Consider $x\notin E$ that is still unqueried. It was not in $D$ by \Cref{lem:conflict}, nor among the singleton points. Thus rounding assigned it through a selected candidate $i\in J$ for some component $J$. By \Cref{lem:inflate}, $x\in B(s_i,3R_i)\subseteq B'_J$. This component must be large, for otherwise its whole old list, including $x$, would have been queried. By \Cref{lem:goodmajority,lem:median}, $x$ passes the test for $J$ and belongs to its recomputed list. Thus only points in $E$ can remain uncovered, and their number is bounded by \eqref{eq:Esize}.
\end{proof}

\begin{lemma}\label{lem:repairqueries}
Whenever rounding does not abort, repair makes $O(kh)=O(k\log n)$ strong queries, even on runs that abort during repair.
\end{lemma}
\begin{proof}
The first abort guard limits the number of conflict queries to $2ckh$. At that point $S_D$ contains at most $2ckh$ points queried during sampling, $ckh$ queried singletons, and $2ckh$ conflicts, so $|S_D|\le5ckh$.

Different unions $B'_J$ are disjoint: if $B(s_i,5R_i)$ and $B(s_j,5R_j)$ intersect, then $d(s_i,s_j)\le5(R_i+R_j)\le7(R_i+R_j)$, placing $i$ and $j$ in the same component. Hence
\[
 \sum_J|S_D\cap B'_J|\le|S_D|,
 \qquad\sum_J|J|\le ck.
\]
All components are classified against the same fixed $S_D$. Their old lists are disjoint, so the total number of small-list queries is at most
\[
 2c\sum_{J\text{ small}}(|S_D\cap B'_J|+|J|h)
 \le2c(|S_D|+ckh)=O(kh).
\]
Large components use at most $ckh$ witness queries. The final guard permits at most $2ckh$ residual queries. All guards precede their queries, so this bound holds on every run, including an aborted run.
\end{proof}

\begin{proof}[Proof of \Cref{lem:repair}]
On the events of \Cref{lem:globalcorrupt,lem:goodmajority}, \Cref{lem:conflict,lem:residual} show that neither abort condition occurs. Small components have empty lists. For a large component, \Cref{lem:goodmajority,lem:median} show that every point passing its test lies in $B(c_J,3R_J)$, proving the first assertion. Any point that is not yet strongly queried and not on a recomputed list is strongly queried in the final step, proving $S\cup\bigcup_JX_J=X$. The query bound follows from \Cref{lem:repairqueries}. A union bound over the concentration events gives high probability for all these conclusions simultaneously.
\end{proof}

\subsection{Compression}\label{sec:compression}

On the event of \Cref{lem:repair}, every point is either strongly queried or belongs to a list contained in $B(c_J,3R_J)$. The number of nonempty lists may still exceed $k$. Compression reduces the solution to at most $k$ balls without further strong queries: solve minimum sum of radii on the exact metric induced by $S$, and enlarge the resulting balls to cover the lists attached to their roots $c_J\in S$. Fix a constant $0<\epsilon\le1$ for the offline approximation below.

\begin{algorithm}[H]
\caption{\textup{\textsc{Compress}}: reducing the number of balls.}\label{alg:compression}
\LinesNumbered
\KwIn{The strong set $S$ and the repaired lists $X_J$ with roots $c_J$ and radii $R_J$}
Run the polynomial-time $(3+\epsilon)$-approximation of \cite{buchem2024msr} on the exact metric restricted to $S$, obtaining at most $k$ balls $\mathcal B$\;
Assign every point of $S$ to a ball of $\mathcal B$ covering it\;
\For{each nonempty list $X_J$}{
 Let $B(z,r)\in\mathcal B$ be the ball to which its root $c_J$ was assigned; assign all points of $X_J$ to that ball\;
}
\For{each $B(z,r)\in\mathcal B$}{
 Increase its radius by $\max\bigl(\{3R_J:\text{$X_J$ was assigned to this ball}\}\cup\{0\}\bigr)$\;
}
\KwRet{The enlarged balls and their point assignment}\;
\end{algorithm}

If a point belongs to several repaired lists, assign it via any one of them. The result is well defined because every list root $c_J$ is strongly queried.

\begin{lemma}\label{lem:restricted}
For every subset $S\subseteq X$, the optimum minimum-sum-of-radii cost on the metric induced by $S$, with centres restricted to $S$, is at most $2\OPT$.
\end{lemma}
\begin{proof}
Take an optimal clustering $X_i^*\subseteq B(o_i,r_i)$. For each $i$ with $X_i^*\cap S\ne\varnothing$, choose $z_i\in X_i^*\cap S$. Every $x\in X_i^*\cap S$ satisfies
\[
 d(z_i,x)\le d(z_i,o_i)+d(o_i,x)\le2r_i.
\]
The resulting at most $k$ balls cover $S$ and have total radius at most $2\sum_i r_i=2\OPT$.
\end{proof}

\begin{lemma}\label{lem:compression}
Fix a guess $\lambda>0$. With high probability, if the run completes sampling, rounding, repair, and compression without aborting, it returns at most $k$ balls and a valid assignment of all points. If $\lambda\ge\OPT/(2k)$, then, with high probability, the run does not abort and its total radius is $O(k\lambda)$.

More precisely, if $\lambda\ge\OPT/(2k)$ and the guarantees of \Cref{lem:accepted,lem:repair} hold, then for $0<\epsilon\le1$ the total radius is less than $2158k\lambda$, so it is at most $Ck\lambda$ for the fixed constant $C:=3000$.
\end{lemma}
\begin{proof}
The offline solution covers $S$, and compression only enlarges its balls. On the event of \Cref{lem:repair}, every point not in $S$ belongs to at least one repaired list. If $x\in X_J$ is assigned through the root $c_J$ to an offline ball $B(z,r)$, then $d(z,c_J)\le r$ and $d(c_J,x)\le3R_J$, so
\[
 d(z,x)\le r+3R_J.
\]
Thus enlarging that offline ball by at least $3R_J$ makes it cover $x$. This proves validity and the bound of $k$ balls whenever the preceding probabilistic guarantees hold and the run does not abort.

Now suppose $\lambda\ge\OPT/(2k)$. By \Cref{lem:rounding,lem:repair}, the run does not abort with high probability. Let $\mathcal B$ be the offline balls before enlargement. By \Cref{lem:restricted},
\[
 \sum_{B(z,r)\in\mathcal B}r
 \le2(3+\epsilon)\OPT.
\]
The total enlargement is at most $3\sum_JR_J$. Hence, by \Cref{lem:components,lem:accepted},
\[
 \begin{aligned}
 \sum_{B\in\mathcal B}\operatorname{radius}(B\text{ after enlargement})
 &\le2(3+\epsilon)\OPT+3\sum_JR_J\\
 &=2(3+\epsilon)\OPT+42\sum_{i=1}^mR_i.
 \end{aligned}
\]
On the events of \Cref{lem:fewcorrupt,lem:sampling}, \Cref{lem:accepted} gives $\sum_iR_i<51k\lambda$. Using $\OPT\le2k\lambda$ and $\epsilon\le1$, the total radius is less than
\[
 4(3+\epsilon)k\lambda+42\cdot51k\lambda
 \le(16+2142)k\lambda=2158k\lambda.
\]
The constant $C=3000$ is therefore sufficient.
\end{proof}

\begin{lemma}\label{lem:runqueries}
A run with any fixed positive guess $\lambda$ makes $O(k\log n)$ strong queries, including if it aborts, and takes polynomial time.
\end{lemma}
\begin{proof}
Sampling makes at most $2ckh=O(k\log n)$ queries. Rounding queries at most $ckh$ singletons because it checks their number first. By \Cref{lem:repairqueries}, repair uses $O(kh)=O(k\log n)$ additional queries; compression uses none. The LP has polynomially many variables and constraints, and each round adds at most $n$ candidate radii per newly sampled centre. All selection, graph, and witness operations are polynomial. Finally, the offline approximation from \cite{buchem2024msr} runs in polynomial time.
\end{proof}

\subsection{Choosing the guess $\lambda$}\label{sec:guessing}

So far, the algorithm assumed a positive guess $\lambda$. For $\lambda\ge\OPT/(2k)$, a run succeeds with high probability and produces a solution of cost at most $Ck\lambda$, where $C=3000$ is the constant from \Cref{lem:compression}. We remove this assumption by testing a geometric sequence of guesses. A run is accepted only if it does not abort and its returned balls have total radius at most $Ck\lambda$.

\begin{theorem}\label{thm:main}
There is a polynomial-time algorithm using $O(k\log n\log\log n)$ strong queries that, with high probability, returns at most $k$ balls and an assignment of every point to a ball containing it, with total radius $O(\OPT)$.
\end{theorem}

If $k\ge n$, return one radius-zero ball per point. Hence assume $k<n$. As in our model, suppose the algorithm knows an upper bound $\Delta\le n^{O(1)}$ on the aspect ratio of the metric. Fix two distinct points $u,v$ independently of the weak distances, strongly query them, and write $d_0=d(u,v)>0$. If $d_{\min}$ and $d_{\max}$ are the minimum and maximum distances between distinct points, then
\begin{equation}\label{eq:d0window}
 \frac{d_0}{\Delta}\le d_{\min}\le d_0\le d_{\max}\le\Delta d_0.
\end{equation}
Set
\begin{equation}\label{eq:bounds}
 \lambda_-:=\frac{d_0}{2Ck\Delta},
 \qquad\lambda_+:=\frac{\Delta d_0}{k}.
\end{equation}
Let $\mathrm{TEST}$ consist of $\lambda_-,2\lambda_-,4\lambda_-,\ldots$ up to the largest such value below $\lambda_+$, together with the endpoint $\lambda_+$. Since
\[
 \frac{\lambda_+}{\lambda_-}=2C\Delta^2=n^{O(1)},
\]
the sequence contains $O(\log n)$ values, with consecutive values differing by a factor at most $2$.

To \emph{test} $\lambda$, run the fixed-guess algorithm (\Cref{sec:overview}) and declare the test to \emph{pass} only if the run does not abort \emph{and} its returned balls have total radius at most $Ck\lambda$. Otherwise the test \emph{fails}. For $\lambda\ge\OPT/(2k)$, \Cref{lem:compression} guarantees a pass with high probability. On the same events, any passing test returns a valid solution, regardless of the guess.

\begin{algorithm}[H]
\caption{\textup{\textsc{Guess}}: searching for a suitable $\lambda$.}\label{alg:guessing}
\LinesNumbered
\KwIn{$X,k$, the aspect-ratio bound $\Delta$, and the weak and strong oracles}
\lIf{$k\ge n$}{\KwRet{Radius-zero balls centred at all points, with the trivial assignment}}
Fix two distinct points $u,v$ independently of the weak distances; strongly query them and compute $d_0$, $\lambda_-$, $\lambda_+$, and $\mathrm{TEST}$\;
\lIf{\textup{\textsc{Test}}$(\lambda_+)$ fails}{\KwRet{The ball $B(u,\Delta d_0)$, assigning all points to it}}
\lIf{\textup{\textsc{Test}}$(\lambda_-)$ passes}{\KwRet{Its solution}}
$\lambda_{\mathrm{lb}}\gets\lambda_-$; $\lambda_{\mathrm{ub}}\gets\lambda_+$ and retain its solution\;
\While{the two bounds are not consecutive in $\mathrm{TEST}$}{
 $\lambda\gets$ a middle value of $\mathrm{TEST}$ strictly between the bounds\;
 \lIf{\textup{\textsc{Test}}$(\lambda)$ passes}{$\lambda_{\mathrm{ub}}\gets\lambda$ and retain its solution}
 \lElse{$\lambda_{\mathrm{lb}}\gets\lambda$}
}
\KwRet{The retained solution at $\lambda_{\mathrm{ub}}$}\;
\end{algorithm}

We retain the solution of each passing test, so the algorithm does not need to rerun the test at the final upper endpoint. The fallback ball is always feasible because $d(u,x)\le d_{\max}\le\Delta d_0$ for every $x$.

\begin{lemma}\label{lem:bounds}
With high probability, the test at $\lambda_+$ passes and the test at $\lambda_-$ fails.
\end{lemma}
\begin{proof}
Condition on the guarantees of \Cref{lem:compression} for the tests at $\lambda_+$ and $\lambda_-$. A union bound gives these guarantees simultaneously with high probability. A single ball of radius $d_{\max}$ covers $X$, so
\[
 \OPT\le d_{\max}\le\Delta d_0=k\lambda_+.
\]
Thus $\lambda_+\ge\OPT/(2k)$, and its test passes with high probability by \Cref{lem:compression} and the choice of $C$.

On the other hand, a ball of radius less than $d_{\min}$ contains only its centre. Since $k<n$, every feasible solution has total radius at least $d_{\min}$; in particular, $\OPT\ge d_{\min}$. A passing test at $\lambda_-$ would, on the event ensuring validity, output a feasible solution of cost at most
\[
 Ck\lambda_-=\frac{d_0}{2\Delta}<d_{\min}\le\OPT,
\]
a contradiction. Therefore this test fails on the same high-probability event.
\end{proof}

\begin{lemma}\label{lem:binsearch}
Binary search makes $O(\log\log n)$ tests on every run. With high probability, it returns a valid solution of cost $O(\OPT)$.
\end{lemma}
\begin{proof}
There are $O(\log n)$ grid values, so binary search uses $O(\log\log n)$ tests. Couple independent random choices for one possible run at each grid value. By \Cref{lem:compression} and a union bound, with high probability its guarantees hold for every grid value simultaneously. Work on this event. By \Cref{lem:bounds}, the initial upper test passes and the initial lower test fails. Every guess at least $\OPT/(2k)$ passes on this event, so any failed guess, including the final $\lambda_{\mathrm{lb}}$, satisfies
\[
 \lambda_{\mathrm{lb}}<\frac{\OPT}{2k}.
\]
The search ends with consecutive grid values, so $\lambda_{\mathrm{ub}}\le2\lambda_{\mathrm{lb}}<\OPT/k$. Because the upper test passed, its retained solution is valid and has radius sum at most
\[
 Ck\lambda_{\mathrm{ub}}<C\OPT.
\]
\end{proof}

\begin{proof}[Proof of \Cref{thm:main}]
There are $O(\log\log n)$ fixed-guess runs. To account for adaptive choices of test values, couple the algorithm with independent randomness for a possible run at every value of the $O(\log n)$-element set $\mathrm{TEST}$. A union bound over all these values ensures that every possible test has its stated guarantees simultaneously, with high probability. On this event \Cref{lem:binsearch} gives a valid $O(\OPT)$ solution. By \Cref{lem:runqueries}, each run uses $O(k\log n)$ strong queries and polynomial time. Including the two initial queries, the total is $O(k\log n\log\log n)$, and the full algorithm is polynomial time.
\end{proof}

\section{Lower bound}\label{sec:lowerbound}

We prove that the $\log n$ factor in our upper bound for the minimum sum of radii cannot be avoided.
Specifically, we prove an $\Omega(k\log n)$ lower bound on the expected number of strong queries for algorithms that succeed with high probability.
We also prove the same bound for constant success probability when $n=k^2$.
These bounds hold even if an algorithm sees every weak distance, makes adaptive strong queries, and outputs centres that it never queried.
We do not require the algorithm to output an assignment, so the bounds also apply to algorithms that do.

The hard instances consist of $K=\Theta(k)$ small clusters, each of size $s=\Theta(\log n)$, together with a padding cluster.
Independently, each small cluster may contain one \emph{outlier}, that is, a point that is far from every other point.
Every approximate solution must choose every outlier as a centre.
However, with some probability, the weak distances within a cluster hide its outlier completely.
We first compute this probability, and then we show that finding the hidden outliers requires many strong queries.
For the high-probability bound, one missed outlier already causes failure.
For constant success probability, we choose the parameters so that many outliers remain hidden.

Recall that each edge is corrupted independently with probability exactly $\delta$.
Throughout this section, $\delta\in(0,\frac12)$ and the approximation factor $\alpha\ge1$ are fixed constants.
The constants in $\Omega(\cdot)$ may depend on $\alpha$, $\delta$ and the required success probability.

\emph{The hard metric:}
Fix integers $K\ge1$ and $s\ge3$ with $Ks\le n-2$, and let $k\in\{2K+1,2K+2\}$.
Set $D:=2\alpha(K+1)$.

\medskip
\noindent\fbox{\begin{minipage}{\dimexpr\linewidth-2\fboxsep-2\fboxrule\relax}
\textbf{Construction of the true metric $(X,d)$}
\begin{enumerate}
 \item Partition $X$ into $K$ \emph{small clusters}
 $B_1,\ldots,B_K$, each containing $s$ points, and a \emph{padding
 cluster} $P$ containing the remaining $n-Ks\ge2$ points.
 \item Independently for each $i\in[K]$, with probability $\frac12$
 choose a uniformly random point $z_i\in B_i$ and designate it the
 \emph{outlier} of $B_i$. Otherwise $B_i$ has no outlier.
 \item For distinct $x,y\in X$, set
 \[
 d(x,y)=
 \begin{cases}
 1 & \text{if $x,y$ are non-outliers belonging to the same cluster},\\
 D & \text{otherwise},
 \end{cases} \\
 \text{and set }d(x,x)=0.
 \]
\end{enumerate}
\end{minipage}}
\medskip

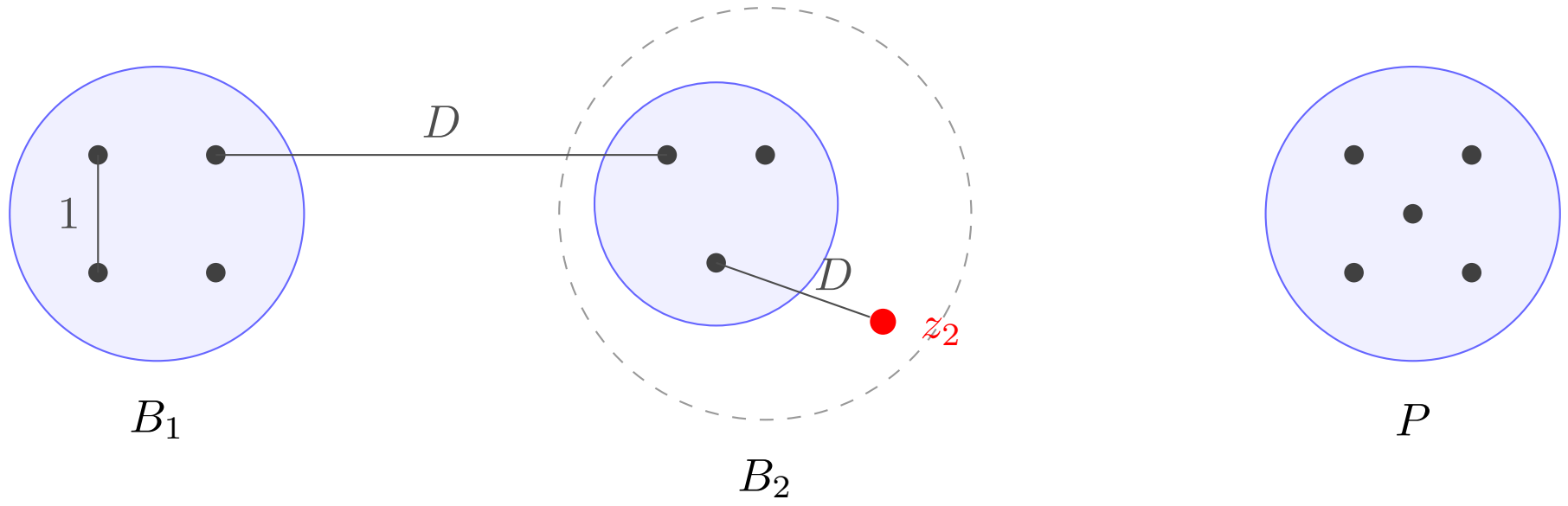
\begin{figure}[ht]
\centering
\begin{tikzpicture}[scale=1.5,
   pt/.style={circle,fill=black!75,inner sep=1.5pt},
   outl/.style={circle,fill=red,inner sep=2pt},
   lbl/.style={font=\small}]
  \draw[blue!60,fill=blue!6] (0,0) circle (0.75);
  \foreach \p in {(-0.3,0.3),(0.3,0.3),(-0.3,-0.3),(0.3,-0.3)} \node[pt] at \p {};
  \node[lbl] at (0,-1.05) {$B_1$};
  \draw[black!40,dashed] (3.1,0) circle (1.05);
  \draw[blue!60,fill=blue!6] (2.85,0.05) circle (0.62);
  \foreach \p in {(2.6,0.3),(3.1,0.3),(2.85,-0.25)} \node[pt] at \p {};
  \node[outl] (z) at (3.7,-0.55) {};
  \node[lbl,text=red,anchor=west] at (3.8,-0.6) {$z_2$};
  \draw[black!70] (z) -- node[lbl,above right,pos=0.4,inner sep=1pt] {$D$} (2.85,-0.25);
  \node[lbl] at (3.1,-1.35) {$B_2$};
  \draw[blue!60,fill=blue!6] (6.4,0) circle (0.75);
  \foreach \p in {(6.1,0.3),(6.7,0.3),(6.1,-0.3),(6.7,-0.3),(6.4,0)} \node[pt] at \p {};
  \node[lbl] at (6.4,-1.05) {$P$};
  \draw[black!70] (0.3,0.3) -- node[lbl,above] {$D$} (2.6,0.3);
  \draw[black!70] (-0.3,0.3) -- node[lbl,left] {$1$} (-0.3,-0.3);
\end{tikzpicture}
\caption{An instance with $K=2$ and $s=4$. Cluster $B_2$ contains
an outlier $z_2$. Distinct non-outliers within one cluster are at
distance $1$; all other pairs are at distance $D$.
The dashed circle marks $B_2$ with its outlier.}
\label{fig:instance}
\end{figure}

This is a metric: if $d(x,y)=d(y,z)=1$, then $x,y,z$ are non-outliers in the same cluster and hence $d(x,z)\le1$; all other triangle inequalities follow since $D\ge1$.
We may give the cluster partition to the algorithm.
The only unknowns are which small clusters contain an outlier and the locations of those outliers (\Cref{fig:instance}).

First, every approximate solution must choose all outliers as centres.

\begin{lemma}\label{lem:centres}
Every $\alpha$-approximate solution for the minimum sum of radii
contains every outlier among its centres.
\end{lemma}

\begin{proof}
Place a radius-$1$ ball at a non-outlier of each of the $K$ small clusters and at a point of $P$, and a radius-$0$ ball at each outlier.
This uses at most $2K+1\le k$ centres and has total radius $K+1$, so $\OPT\le K+1$.
In an $\alpha$-approximate solution, each ball has radius at most $\alpha(K+1)=D/2<D$.
Because an outlier is at distance $D$ from every other point, only a ball centred at the outlier can cover it.
\end{proof}

\emph{What the weak distances reveal:}
We next specify an admissible corruption rule.
Its purpose is to make an outlier indistinguishable from an ordinary point whenever all its incident edges within its cluster are corrupted.

\medskip
\noindent\fbox{\begin{minipage}{\dimexpr\linewidth-2\fboxsep-2\fboxrule\relax}
\textbf{Construction of the weak distances $\widetilde d$}
\begin{enumerate}
 \item Mark every edge as corrupted independently with probability
 $\delta$.
 \item For a corrupted edge whose endpoints lie in the same small
 cluster, set $\widetilde d(x,y)=1$. On every other edge, set
 $\widetilde d(x,y)=d(x,y)$.
\end{enumerate}
\end{minipage}}
\medskip

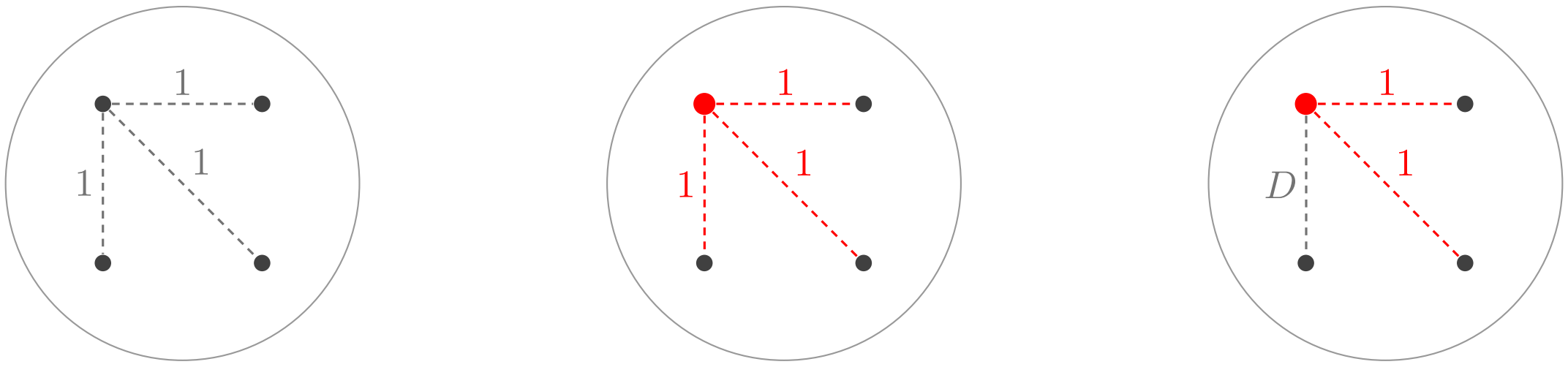
\begin{figure}[ht]
\centering
\begin{tikzpicture}[scale=1.6,
   pt/.style={circle,fill=black!75,inner sep=1.5pt},
   outl/.style={circle,fill=red,inner sep=2pt},
   ok/.style={black!55,dash pattern=on 2pt off 1.6pt,line width=0.6pt},
   bad/.style={red,dash pattern=on 2pt off 1.6pt,line width=0.6pt},
   lbl/.style={font=\footnotesize},
   num/.style={font=\small,inner sep=2pt}]
  \draw[black!40] (0,0) circle (1.0);
  \node[pt] (a0) at (-0.45,0.45) {};
  \node[pt] (a1) at (0.45,0.45) {};
  \node[pt] (a2) at (-0.45,-0.45) {};
  \node[pt] (a3) at (0.45,-0.45) {};
  \foreach \t/\p in {a1/above,a2/left,a3/above right} \draw[ok] (a0) -- node[num,\p] {$1$} (\t);
  \node[lbl,align=center] at (0,-1.45) {no outlier\\ (blank)};
  \draw[black!40] (3.4,0) circle (1.0);
  \node[outl] (b0) at (2.95,0.45) {};
  \node[pt] (b1) at (3.85,0.45) {};
  \node[pt] (b2) at (2.95,-0.45) {};
  \node[pt] (b3) at (3.85,-0.45) {};
  \foreach \t/\p in {b1/above,b2/left,b3/above right} \draw[bad] (b0) -- node[num,\p] {$1$} (\t);
  \node[lbl,align=center] at (3.4,-1.45) {hidden outlier\\ (blank)};
  \draw[black!40] (6.8,0) circle (1.0);
  \node[outl] (c0) at (6.35,0.45) {};
  \node[pt] (c1) at (7.25,0.45) {};
  \node[pt] (c2) at (6.35,-0.45) {};
  \node[pt] (c3) at (7.25,-0.45) {};
  \draw[bad] (c0) -- node[num,above] {$1$} (c1);
  \draw[ok] (c0) -- node[num,left] {$D$} (c2);
  \draw[bad] (c0) -- node[num,above right] {$1$} (c3);
  \node[lbl,align=center] at (6.8,-1.45) {visible outlier\\ (not blank)};
\end{tikzpicture}
\caption{Weak distances from one point in a small cluster with $s=4$
(dashed edges; red edges are corrupted). The first two clusters are
indistinguishable from their weak distances, whereas a correct edge
of length $D$ reveals the presence of an outlier in the third.
The red point is the outlier.}
\label{fig:weak}
\end{figure}

We call a small cluster \emph{blank} if all its internal weak distances are $1$.
A cluster with no outlier is always blank.
A cluster with an outlier is blank exactly when all $s-1$ edges from the outlier to the rest of the cluster are corrupted, and this happens with probability $p:=\delta^{s-1}$.
In both cases, a blank cluster has exactly the same weak distances (\Cref{fig:weak}).

The following lemma gives the posterior distribution even when the algorithm sees \emph{all} weak distances.

\begin{lemma}\label{lem:posterior}
Conditioned on all weak distances, the small clusters remain
independent. For any blank cluster $B_i$, the conditional probability
that it has no outlier is $1/(1+p)$, while each point $x\in B_i$ is
its outlier with conditional probability $p/((1+p)s)$.
\end{lemma}

\begin{proof}
A cluster with no outlier is blank with probability $1$, and a cluster with an outlier is blank with probability $p$.
The prior probability of an outlier is $1/2$, so Bayes' rule gives
\[
 \Pr[B_i\text{ has an outlier}\mid B_i\text{ is blank}]
 =\frac{p/2}{1/2+p/2}=\frac{p}{1+p}.
\]
Conditional on this event, the outlier is uniform over the $s$ points by symmetry, which gives the claimed probabilities.

Every weak distance between different clusters is $D$, and every weak distance within $P$ is $1$.
These weak distances give no information about the outliers.
Within a small cluster, the weak distances depend only on the outlier choice and the edge corruptions of that cluster, and these are independent of the other clusters.
Hence, the conditional distributions of different small clusters are independent.
\end{proof}

\emph{How much a strong query can discover:}
For the analysis, we use an auxiliary \emph{point-probe oracle}.
Probing a point reveals whether it is an outlier.
We can simulate a strong point query at $x$ by probing $x$.
This is because we already know which previously queried points are outliers, and this together with the clusters gives all distances that the query reveals.
We can simulate a strong edge query $(x,y)$ by probing both endpoints.
So an algorithm that makes $Q$ strong queries of either kind can be simulated with at most $2Q$ point probes.
The factor of two does not affect the lower bound.

A \emph{hidden outlier} is an outlier in a blank cluster.
We say it is \emph{found} when it is probed for the first time.
The following bound holds for any adaptive sequence of point probes.

\begin{lemma}\label{lem:rate}
Under the random construction above, let $T$ be the number of point probes of an adaptive algorithm, and let $F$ be the number of hidden outliers it finds.
Then
\[
 \mathbb{E}[F]\le\frac{p}{s}\,\mathbb{E}[T].
\]
\end{lemma}

\begin{proof}
We may reveal all weak distances before the probes, because this information only makes the algorithm stronger.
Fix these weak distances and the coins of the algorithm, and consider its next probe $x\in B_i$.
The probe can find a hidden outlier only if $B_i$ is blank, no outlier has been found in $B_i$ yet, and $x$ has not been probed before.
Suppose $a<s$ distinct points of $B_i$ were probed before, and none of them was an outlier.

By \Cref{lem:posterior}, and because probes in other clusters give no information about $B_i$, the conditional probability that $x$ is the outlier is
\[
 \frac{p/((1+p)s)}{1-a\,p/((1+p)s)}
 =\frac{p}{(1+p)s-ap}
 \le\frac{p}{s}.
\]
The inequality holds because $a<s$.
If the conditions above do not hold, the probability of finding a new hidden outlier is zero.
Therefore, conditional on the history before it, every point probe finds a hidden outlier with probability at most $p/s$.

For each $t\ge1$, let $I_t$ be the indicator that the $t$-th probe finds a hidden outlier, with $I_t=0$ if the algorithm makes fewer than $t$ probes.
Then $F=\sum_{t\ge1} I_t$.
Let $\mathcal H_{t-1}$ be all information that the algorithm has before it sees the answer to its $t$-th probe, including the answers to previous probes and its coins.
In particular, this information decides whether the algorithm makes the $t$-th probe and which point it probes.

By the bound above, conditional on any such history, the next probe finds a hidden outlier with probability at most $p/s$.
That is,
\[
 \mathbb E[I_t\mid\mathcal H_{t-1}] \le \frac{p}{s}\,\mathbf 1_{\{T\ge t\}}.
\]
Taking expectations gives $\mathbb E[I_t] \le \frac{p}{s}\Pr[T\ge t]$.
Finally, by linearity of expectation,
\[
\begin{aligned}
\mathbb E[F]
=\sum_{t\ge1}\mathbb E[I_t]
\le\frac{p}{s}\sum_{t\ge1}\Pr[T\ge t]
=\frac{p}{s}\,\mathbb E[T],
\end{aligned}
\]
where the last equality is the tail-sum formula for nonnegative integer-valued random variables.

\end{proof}

We next connect finding hidden outliers to a successful output, also when the algorithm may output unqueried centres.

\begin{lemma}\label{lem:output-probes}
Let an algorithm make $Q$ strong point or strong edge queries and output at most $k$ centres.
We can simulate its execution and postprocess its output with at most $T\le 2Q+k$ point probes, so that every hidden outlier is found whenever its output is $\alpha$-approximate.
Hence, if $F$ is the number of hidden outliers found after postprocessing, then $\mathbb{E}[F]\le\frac{p}{s}\bigl(2\mathbb{E}[Q]+k\bigr)$.
\end{lemma}

\begin{proof}
We simulate each strong point query with one point probe, and each strong edge query with at most two point probes.
After the algorithm outputs its centres, we probe each output centre that was not probed before, which uses at most $k$ more probes.
If the output is $\alpha$-approximate, then every outlier is an output centre by \Cref{lem:centres}, and all hidden outliers are found.
The bound on $\mathbb{E}[F]$ follows from \Cref{lem:rate}.
\end{proof}

Let $M$ be the total number of hidden outliers.
A small cluster contains a hidden outlier with probability $p/2$, independently of the other clusters.
Thus
\begin{equation}\label{eq:hidden-moments}
 M\sim\operatorname{Bin}(K,p/2),\qquad
 \mu:=\mathbb{E}[M]=\frac{Kp}{2},\qquad
 \operatorname{Var}(M)\le\mu,
 \qquad\mathbb{E}[M^2]\le\mu+\mu^2.
\end{equation}
We now apply these facts with two choices of $s$.

\emph{High-probability success:}
Here, missing a single hidden outlier is already enough to break the required failure probability.

\begin{theorem}\label{thm:whp}
Fix $\beta>0$. For all sufficiently large $n$ and
$3\le k\le n/\log^2 n$, every algorithm that returns an
$\alpha$-approximate minimum-sum-of-radii solution with probability at
least $1-n^{-\beta}$ on every instance makes
$\Omega(k\log n)$ strong queries in expectation on some instance.
\end{theorem}

\begin{proof}
Choose
\[
 K:=\left\lfloor\frac{k-1}{2}\right\rfloor,
 \qquad
 s:=1+\left\lfloor\frac{\beta}{2}\log_{1/\delta} n\right\rfloor.
\]
Then $K=\Theta(k)$, $k\in\{2K+1,2K+2\}$, $k\le4K$, and $s=\Theta(\log n)$.
For large enough $n$, we have $s\ge32$ and $Ks\le n-2$, so the construction is valid.
Moreover,
\[
 p=\delta^{s-1}\ge n^{-\beta/2},
 \qquad \mu=\frac{Kp}{2}\ge\frac12n^{-\beta/2}.
\]

All expectations and probabilities below are over the random metric, the edge corruptions and the coins of the algorithm.
Suppose, for a contradiction, that $\mathbb{E}[Q]\le Ks/16$.
We apply \Cref{lem:output-probes}.
Because $k\le4K\le Ks/8$,
\[
 \mathbb{E}[T]\le2\mathbb{E}[Q]+k
 \le\frac{Ks}{4},
 \qquad
 \mathbb{E}[F]\le\frac{p}{s}\,\mathbb{E}[T]
 \le\frac{Kp}{4}=\frac\mu2.
\]
Let $Y:=M-F$ be the number of hidden outliers that are not found.
Then
\[
 \mathbb{E}[Y]\ge\frac\mu2,
 \qquad 0\le Y\le M.
\]
Because $Y\ge 0$, the Cauchy--Schwarz inequality gives
\[
 \mathbb E[Y] =\mathbb E[Y\mathbf 1_{\{Y>0\}}] \le \sqrt{\mathbb E[Y^2]\Pr[Y>0]}.
\]
With \eqref{eq:hidden-moments}, this gives
\[
\Pr[Y>0]
\ge\frac{\mathbb E[Y]^2}{\mathbb E[Y^2]}
\ge\frac{\mu^2/4}{\mu+\mu^2}
=\frac{\mu}{4(1+\mu)}
\ge\frac{\min\{\mu,1\}}8
\ge\frac1{16}n^{-\beta/2}.
\]
For large enough $n$, this is more than $n^{-\beta}$.
Whenever $Y>0$, the output of the algorithm is not $\alpha$-approximate by \Cref{lem:output-probes}.
So the average failure probability of the algorithm under the random construction is more than the permitted value.
Then some fixed true metric in the support violates the success guarantee, which is a contradiction.

We conclude that $\mathbb{E}[Q]>Ks/16$ under the random construction.
Averaging over the true metrics, some fixed metric requires more than $Ks/16=\Omega(k\log n)$ strong queries in expectation over the corruptions and the coins of the algorithm.
\end{proof}

\emph{Constant success probability:}
For constant success probability, we choose $s$ so that the expected number $\mu$ of hidden outliers grows with $k$.
Then a successful algorithm must find many of them.

\begin{theorem}\label{thm:constant}
Fix $\rho\in(0,1]$. For all sufficiently large $k$, let $n=k^2$.
Every algorithm that returns an $\alpha$-approximate
minimum-sum-of-radii solution with probability at least $\rho$ on
every instance makes $\Omega(k\log n)$ strong queries in expectation
on some instance.
\end{theorem}

\begin{proof}
Set
\[
 K:=\left\lfloor\frac{k-1}{2}\right\rfloor,
 \qquad
 s:=1+\left\lfloor\frac12\log_{1/\delta}K\right\rfloor.
\]
For large enough $k$, we have $s\ge3$ and $Ks=O(k\log k)\le n-2$.
Moreover,
\[
 p=\delta^{s-1}\ge K^{-1/2},
 \qquad
 \mu=\frac{Kp}{2}\ge\frac12\sqrt K\longrightarrow\infty.
\]

Let $S$ be the event that the algorithm succeeds, so $\Pr[S]\ge\rho$ over the random instance.
We postprocess its output as in \Cref{lem:output-probes}.
On $S$, all $M$ hidden outliers are found, and hence $F=M$.
Using \eqref{eq:hidden-moments},
\[
 \begin{aligned}
 \mathbb{E}[F]
 \ge\mathbb{E}[M\mathbf{1}_S]
 \ge\mu\Pr[S]-\mathbb{E}[|M-\mu|]
 \ge\rho\mu-\sqrt{\operatorname{Var}(M)}
 \ge\rho\mu-\sqrt\mu.
 \end{aligned}
\]
With \Cref{lem:output-probes}, this gives
\[
 \mathbb{E}[Q] \ge\frac{s}{2p}\bigl(\rho\mu-\sqrt\mu\bigr)-\frac k2 =\left(\frac\rho4-\frac1{4\sqrt\mu}\right)Ks-\frac k2.
\]
Because $\mu\to\infty$, $s=\Theta(\log k)$ and $k=\Theta(K)$, the right-hand side is $\Omega(Ks)=\Omega(k\log k)$.
As $n=k^2$, this is $\Omega(k\log n)$.
Averaging over the true metrics again gives a fixed metric with the stated lower bound on the expected number of queries.
\end{proof}

For fixed $\rho$ and all large enough $k$, the same proof gives $\Omega(k\log k)$ strong queries in expectation for every $n\ge k\log_{1/\delta}k$, if we place the extra points in $P$.

\emph{Extension to $k$-center:}
The only property of the minimum sum of radii that we used is \Cref{lem:centres}.
So the same construction gives the same bounds for $k$-center.

\begin{corollary}\label{cor:kcenter}
\Cref{thm:whp,thm:constant} also hold for $k$-center when the
algorithm outputs only a set of centres.
\end{corollary}

\begin{proof}
We replace \Cref{lem:centres} with its $k$-center version.
One radius-$1$ ball at a non-outlier of each cluster and a radius-$0$ ball at every outlier give $k$-center cost at most $1$.
In the parameter ranges of both theorems, $k<n$, and every two distinct points are at distance at least $1$.
Hence $k$ balls of radius less than $1$ cannot cover all $n$ points, and the optimum is exactly $1$.
An $\alpha$-approximate solution has covering radius at most $\alpha<D$, so it must choose every outlier as a centre.
The rest of the argument is unchanged.
\end{proof}

\section{Conclusion}\label{sec:conclusion}

We gave a polynomial-time $O(1)$-approximation for MSR with $O(k\log n\log\log n)$ strong queries, and we showed that $\Omega(k\log n)$ strong queries are necessary for algorithms that succeed with high probability.
The $O(\log\log n)$ factor comes only from the binary search, and removing it is an open problem.
For $k$-median and $k$-means the best known bound is still $O(k\log^2 n)$ strong queries, and we do not know if a packing LP can improve it.

\section*{AI methodology}

We used Gemini (Google) and GPT Astra  to assist with this paper.
The use was as follows.
\begin{itemize}[leftmargin=*]
 \item \emph{Proofs:} Gemini 3.1 Pro was used to verify proofs of elementary probability lemmas \ref{lem:fewcorrupt}, \ref{lem:round}, \ref{lem:uniform}, and \ref{lem:honest}
 \item \emph{Research:} The hardness construction was developed via the authors' interaction with GPT 6. 
\end{itemize}
The algorithms, results and proofs are authors' own, and they checked every proof and every statement in the paper.
The authors take full responsibility for its content.

\bibliographystyle{alpha}
\bibliography{arxiv/ref}

\appendix
\section{General corruption probability}\label{app:delta}

Fix $\delta\in(0,\frac12)$ and let $\gamma:=\frac12-\delta$.
The algorithm knows $\delta$.
We change two parameters.
Let
\begin{equation}\label{eq:hgamma}
 h:=32\left\lceil\frac{c\log n}{32\gamma^2}\right\rceil,
\end{equation}
and call a component $J$ small when
\begin{equation}\label{eq:smallgamma}
 |X_J|<\frac{2c}{\gamma}\left(|S_D\cap B'_J|+|J|h\right).
\end{equation}
As in the main algorithm, we classify all components with the same set $S_D$.
Everything else is unchanged, including the distance tests, the radius factors and the two repair guards $|D|\le2ckh$ and $|U|\le2ckh$.
As before, $c$ is a large enough absolute constant, and it does not depend on $\delta$.

\begin{theorem}\label{thm:generaldelta}
For every fixed $\delta\in(0,\frac12)$, the algorithm with these parameters makes
\[
 O\!\left(\frac{k\log n\log\log n}{(\tfrac12-\delta)^3}\right)
\]
strong point queries and runs in polynomial time.
With high probability, it returns at most $k$ covering balls and an assignment of every point to a ball containing it, of total radius $O(\OPT)$.
The approximation constant is independent of $\delta$.
\end{theorem}

\begin{proof}
The main text uses $\delta=1/3$ only in concentration bounds.
Each of them needs the fraction of corrupted edges to stay a constant below $\frac12$.
With margin $\gamma$, each bound fails with probability $e^{-\Omega(\gamma^2h)}$, and by \eqref{eq:hgamma} this is again $n^{-\Omega(c)}$.
We list the bounds that change.

\emph{Corrupt points:}
For a fixed $h$-point set $W$ and $x\notin W$, \Cref{lem:hoeffding} gives
\[
 \Pr[x\text{ is corrupt for }W]\le e^{-2\gamma^2h}.
\]
With this bound, the union bounds in the proofs of \Cref{lem:fewcorrupt,lem:globalcorrupt} fail with probability at most
\[
 n^{2h}e^{-\gamma^2h^2}\le n^{-(c-2)h}
 \qquad\text{and}\qquad
 n^{2ckh}(ckh\,e^{-2\gamma^2h})^{ckh}\le n^{-(2c-3)ckh}.
\]
The sampling and rounding proofs use $\delta$ only through these two lemmas, so they are unchanged.

\emph{Repair:}
Let $a:=\frac12-3\gamma/4$.
For a point $x$ and a union $A$ of at most $j$ true balls with $|A|\ge cjh$, \Cref{lem:hoeffding} gives
\[
 \Pr[\text{more than }a|A|\text{ edges from }x\text{ to }A\setminus\{x\}\text{ are corrupted}]
 \le e^{-\gamma^2|A|/8}\le n^{-c^2j/8}.
\]
A union bound over the $n^{2j}$ unions and $n$ points, as in \Cref{lem:uniform}, makes this hold for all such $A$ and $x$.

Fix a large component $J$, and assume the event above that replaces \Cref{lem:fewcorrupt}.
By \Cref{lem:listfast} and \eqref{eq:smallgamma},
\[
 |X_J\setminus B'_J|<\frac{|J|h}{2}\le\frac{\gamma}{4c}|X_J|
 \qquad\text{and}\qquad
 |S_D\cap B'_J|+|E_J|<\frac{\gamma}{2c}|X_J|.
\]
Also,
\[
 |B'_J|\ge|X_J|-|E_J|>\left(\frac{2c}{\gamma}-\frac12\right)|J|h\ge c|J|h,
\]
so the bound above applies to $B'_J$.
Then the fraction of points of $X_J$ whose edge to a fixed point $x$ is corrupted is at most
\[
 a+(a+1)\frac{\gamma}{2c}\le\frac12-\frac{5\gamma}{8},
\]
where we use $c\ge6$.
Conditional on the old list and the set $\mathsf{Corrupt}$, the witnesses $W_J$ are a fresh uniform sample from $X_J$.
By \Cref{lem:hypergeom} and a union bound over $ck$ components and $n$ points,
\[
 \Pr[|W_J\setminus B'_J|\ge\gamma h/4]\le(e/c)^{\gamma h/4},
\]
\[
 \Pr[\text{at least }(\tfrac12-\tfrac\gamma2)h\text{ witness edges from }x\text{ are corrupted}]\le e^{-\gamma^2h/32}.
\]
The first bound is $n^{-\Omega(c)}$ because $\gamma h\ge2c\log n$, and the second is at most $n^{-c/32}$ because $\gamma^2h\ge c\log n$.
Hence every tested point has more than $h/2$ witnesses that are honest and inside $B'_J$.
\Cref{lem:median,lem:residual} follow with the same distance factors.

\emph{Queries and cost:}
Only the number of small-list queries changes.
The unions $B'_J$ are disjoint, $|S_D|\le5ckh$ and $\sum_J|J|\le ck$, so this number is at most
\[
 \frac{2c}{\gamma}\sum_{J\text{ small}}(|S_D\cap B'_J|+|J|h)
 \le\frac{2c}{\gamma}(|S_D|+ckh)=O(kh/\gamma).
\]
All other counts are $O(kh)$.
A run uses $O(kh/\gamma)=O(k\log n/\gamma^3)$ strong queries, and the search makes $O(\log\log n)$ runs.
The geometric bounds are unchanged, so the approximation constant does not depend on $\delta$, and the running time is polynomial.
\end{proof}

\section{Concentration bounds}\label{app:tools}

We state the three bounds that we use, in the form we use them.
The first is \emph{Hoeffding's inequality}~\cite{hoeffding1963}.

\begin{lemma}\label{lem:hoeffding}
Let $X_1,\dots,X_m$ be independent random variables with values in $[0,1]$, and let $X:=\sum_{i=1}^mX_i$.
Then for every $t>0$,
\begin{equation*}
 \Pr\left[X\ge\mathbb{E}[X]+t\right]\le e^{-2t^2/m}.
\end{equation*}
\end{lemma}

We often apply \Cref{lem:hoeffding} with an upper bound $\mu\ge\mathbb{E}[X]$ in place of $\mathbb{E}[X]$.
This is allowed because
\begin{equation*}
 \Pr[X\ge\mu+t]\le\Pr\left[X\ge\mathbb{E}[X]+t\right]\le e^{-2t^2/m}.
\end{equation*}
It is used in \Cref{lem:fewcorrupt,lem:globalcorrupt,lem:uniform}.

The second is the \emph{multiplicative Chernoff bound}, in the form of Mitzenmacher and Upfal~\cite[Theorems~4.4 and~4.5]{mitzenmacher2017probability}.

\begin{lemma}\label{lem:chernoff}
Let $X_1,\dots,X_m$ be independent random variables with values in $[0,1]$, and let $X:=\sum_{i=1}^mX_i$.
\begin{enumerate}
 \item If $\mathbb{E}[X]\le\mu$ and $a>\mu$, then
 \begin{equation*}
  \Pr[X\ge a]\le\left(\frac{e\mu}{a}\right)^a.
 \end{equation*}
 \item If $\mathbb{E}[X]\ge\mu>0$ and $0<\epsilon<1$, then
 \begin{equation*}
  \Pr\left[X\le(1-\epsilon)\mu\right]\le e^{-\epsilon^2\mu/2}.
 \end{equation*}
\end{enumerate}
\end{lemma}

The first part follows from the standard upper tail.
For $\mathbb{E}[X]\le\mu$ and $a=(1+\epsilon)\mu$ with $\epsilon>0$, it reads
\begin{equation*}
 \Pr[X\ge a]\le\left(\frac{e^{\epsilon}}{(1+\epsilon)^{1+\epsilon}}\right)^{\mu}=e^{a-\mu}\left(\frac{\mu}{a}\right)^{a}\le\left(\frac{e\mu}{a}\right)^{a}.
\end{equation*}
The first part is used through \Cref{lem:hypergeom}, and the second part in \Cref{lem:round} with $\epsilon=\frac12$.

The third is the \emph{hypergeometric Chernoff bound}, which follows from Hoeffding's comparison of sampling without and with replacement~\cite[Theorem~4]{hoeffding1963}.

\begin{lemma}\label{lem:hypergeom}
Let $P$ be a finite set in which some points are marked, and let $m\le|P|$.
Let $X$ be the number of marked points among $m$ points drawn uniformly without replacement from $P$, and let $\mathbb{E}[X]\le\mu$.
\begin{enumerate}
 \item For every $t>0$,
 \begin{equation*}
  \Pr[X\ge\mu+t]\le e^{-2t^2/m}.
 \end{equation*}
 \item For every $a>\mu$,
 \begin{equation*}
  \Pr[X\ge a]\le\left(\frac{e\mu}{a}\right)^a.
 \end{equation*}
\end{enumerate}
\end{lemma}
\begin{proof}
Let $Y$ be the number of marked points among $m$ points drawn uniformly and independently with replacement.
Then $\mathbb{E}[X]=\mathbb{E}[Y]$, and $Y$ is a sum of independent random variables.
Theorem~4 of Hoeffding~\cite{hoeffding1963} gives $\mathbb{E}[f(X)]\le\mathbb{E}[f(Y)]$ for every continuous convex function $f$.
The proofs of \Cref{lem:hoeffding,lem:chernoff} bound each probability through $\mathbb{E}[e^{sY}]$ for a real $s$, and $x\mapsto e^{sx}$ is continuous and convex.
So both bounds hold for $X$ as they hold for $Y$.
\end{proof}

The first part is used in \Cref{lem:honest}, where a point is marked when its edge to $x$ is corrupted.
The second part is used in \Cref{lem:inside}, where a point is marked when it lies outside $B'_J$.
In both cases $P$ is the list $X_J$.

\section{The approximation factor}\label{app:factor}

The analysis in \Cref{sec:guessing} uses the factor $C=3000$ only in the test and in $\lambda_-$.
Here we compute the approximation factor for one choice of the parameters.

\begin{theorem}\label{thm:factor}
For every constant $\delta\in(0,\frac12)$, $k<n$, offline parameter $\epsilon=1/100$ and a constant $c\ge20000$, with high probability the algorithm of \Cref{thm:main}, run with the parameters of \Cref{app:delta}, returns a solution of total radius less than $270\,\OPT$.
\end{theorem}
\begin{proof}
Let $\xi=1/1000$.
We change only the analysis of sampling, and we call cluster $i$ heavy if
\[
 y(F_i)>6r_i+\tfrac32\lambda.
\]

\emph{A captured cluster is not heavy:}
The capture proof takes the largest estimate from the cluster as the radius $R\le6r_i$, and this candidate has weight more than $R+\lambda$.

\emph{A cluster receives fewer than $7h$ draws while it is heavy:}
Every point has weight at most $\lambda/h$, and a heavy cluster has weight more than $3\lambda/2$.
So while at most $h$ distinct points of $X_i^*$ have been drawn, a draw into $F_i$ in a heavy round repeats one of them with probability less than
\[
 h\cdot\frac{\lambda/h}{3\lambda/2}=\frac23.
\]
Then each of the first $6h$ such draws gives a new point with probability at least $1/3$, until $h+1$ points are drawn.
By \Cref{lem:chernoff} with $\mu=2h$ and $\epsilon=\frac12$, which holds for such sequential draws because each succeeds with probability at least $1/3$ given the earlier ones,
\[
 \Pr[\text{the first }6h\text{ draws into }F_i\text{ contain at most }h\text{ distinct points}]\le e^{-h/4}.
\]
By \Cref{lem:capture}, the round that draws the $(h+1)$-th distinct point captures the cluster.
This round adds at most $h$ draws, so the cluster receives fewer than $7h$ draws while it is heavy.

\emph{LP bound:}
Let $p=\xi/(1+\xi)$, and consider a round whose LP optimum is at least $(1+\xi)(6\OPT+\frac32k\lambda)$.
The argument of \Cref{lem:round} gives
\[
 \Pr[\text{the round has fewer than }ph/2\text{ draws into heavy clusters}]\le e^{-ph/8}.
\]
For $c\ge20000$ we have $ph/8\ge2.49\log n$, so a union bound over the $ck$ rounds and the $O(\log n)$ guesses still fails with probability $n^{-\Omega(1)}$.
If all $ck$ rounds had such an LP optimum, the heavy clusters would receive at least $ck\cdot ph/2\ge7kh$ draws, because $c>14/p$.
This is not possible. Hence, with high probability,
\[
 \OPTLP<(1+\xi)\left(6\OPT+\tfrac32k\lambda\right).
\]

\emph{Cost:}
The parts of the proofs of \Cref{lem:accepted,lem:compression} that do not use $\lambda\ge\OPT/(2k)$ give
\[
 \sum_i\left(R_i+\frac\lambda2\right)<\OPTLP
 \qquad\text{and}\qquad
 F\le2(3+\epsilon)\OPT+42\sum_iR_i.
\]
With the LP bound, every run that completes has
\[
 F<2(3+\epsilon)\OPT+42(1+\xi)\left(6\OPT+\tfrac32k\lambda\right)=A\,\OPT+Bk\lambda,
\]
where $A=258.272$ and $B=63.063$.

\emph{Search:}
Let $t=A/(3000-B)$, and consider a guess $\lambda\ge t\OPT/k$.
Rounding does not abort, because the LP bound gives
\[
 |L|\le\frac h\lambda\OPTLP<(1+\xi)\left(\frac6t+\frac32\right)kh<70kh
 \qquad\text{and}\qquad
 |Z'|<\frac2\lambda\OPTLP<140k.
\]
The guards of repair do not depend on $\lambda$.
By the cost bound, $F<A\,\OPT+Bk\lambda\le3000k\lambda$, and this guess passes the test.
Hence every failed guess is less than $t\OPT/k$.
The final upper guess is at most twice the last failed guess, which gives $\lambda_{\mathrm{ub}}<2t\OPT/k$.
The cost bound for the retained solution gives
\[
 F<(A+2Bt)\OPT=A\cdot\frac{3000+B}{3000-B}\cdot\OPT<270\,\OPT.
\]
All other parameters are constants, so the running time is polynomial and the number of strong queries is as in \Cref{thm:generaldelta}.
\end{proof}

\end{document}